\documentclass[aps,prc,superscriptaddress,nofootinbib,showpacs,floatfix,singlecolumn]{revtex4}
\usepackage{url}
\usepackage{cancel}
\usepackage[colorlinks,linkcolor=blue,citecolor=blue,filecolor=black,urlcolor=blue]{hyperref}
\usepackage{epsfig,graphics}
\usepackage{graphicx}
\usepackage{dcolumn}
\usepackage{bm}
\usepackage[usenames]{color}
\usepackage{amssymb}
\usepackage{amsmath}
\usepackage{multirow}
\usepackage{float}
\usepackage{harpoon}
\usepackage{MnSymbol}
\usepackage{appendix}
\usepackage{color}
\usepackage{hyperref}
\usepackage{cleveref}

\newcommand{\sqrtsnn}{\mbox{$\sqrt{s_{\mathrm{NN}}}$}}
\newcommand{\pT} {p_{\mathrm{T}}}
\newcommand{\lr}[1]{\left\langle #1\right\rangle}

\newcommand{\nch}{N_{\mathrm{ch}}}

\newcommand{\npart}{N_{\mathrm{part}}}
\newcommand{\nqp}{N_{\mathrm{quark}}}

\usepackage{CJK}
\hfuzz=\maxdimen
\begin{document}
\title{Shape of atomic nuclei in heavy ion collisions}
\newcommand{\sbu}{Department of Chemistry, Stony Brook University, Stony Brook, NY 11794, USA}
\newcommand{\bnl}{Physics Department, Brookhaven National Laboratory, Upton, NY 11976, USA}
\author{Jiangyong Jia}\email[Correspond to\ ]{jiangyong.jia@stonybrook.edu}\affiliation{\sbu}\affiliation{\bnl}
\date{\today}
\begin{abstract}
In the hydrodynamic model description of heavy ion collisions, the final-state anisotropic flow $v_n$ are linearly related to the strength of the multi-pole shape of the distribution of nucleons in the transverse plane $\varepsilon_n$, $v_n\propto \varepsilon_n$. The $\varepsilon_n$, for $n=1,2,3,4$, are sensitive to the shape of the colliding ions, characterized by the quadrupole $\beta_2$, octupole $\beta_3$ and hexadecapole $\beta_4$ deformations. This sensitivity is investigated analytically and also in a Monte Carlo Glauber model. One observes a robust linear relation, $\langle\varepsilon_n^2\rangle = a_n'+b_n'\beta_n^2$, for events in a fixed centrality. The $\langle\varepsilon_1^2\rangle$ has a contribution from $\beta_3$ and $\beta_4$, and $\langle\varepsilon_3^2\rangle$ from $\beta_4$. In the ultra-central collisions, there are little cross contributions between $\beta_2$ and $\varepsilon_3$ and between $\beta_3$ and $\varepsilon_2$, but clear cross contributions are present in non-central collisions. Additionally, $\langle\varepsilon_n^2\rangle$ are insensitive to non-axial shape parameters such as the triaxiality. This is good news because the measurements of $v_2$, $v_3$ and $v_4$ can be used to constrain simultaneously the $\beta_2$, $\beta_3$, and $\beta_4$ values. This is best done by comparing two colliding ions with similar mass numbers and therefore nearly identical $a_n'$, to obtain simple equation that relates the $\beta_n$ of the two species. This opens up the possibility to map the shape of the atomic nuclei at a timescale ($<10^{-24}$s) much shorter than probed by low-energy nuclear structure physics ($<10^{-21}$s), which ultimately may provide information complementary to those obtained in the nuclear structure experiments.
\end{abstract}

\maketitle
\tableofcontents
\clearpage
\section{Introduction}\label{sec:1}
Most of the  atomic nuclei in their ground state are deformed from a well defined spherical shape. The deformation has non-trivial dependence on the proton and neutron number, especially in the vicinity of full shell or subshell, reflecting collective motion induced by interaction between valence nucleons and shell structure~\cite{bohr}. The collective motion leads to characteristic rotational spectra of nuclear excited state, where the electric multi-pole transition probability $B(En)$ between low-lying rotational states with $n\hbar$ difference in angular momentum can be used to infer the shape parameters. Past efforts have led to the discovery of a rich variety of phenomena, such as quadrupole deformation, shape evolution, triaxiality/shape coexistence, octupole deformation, hexadecapole deformation and other exotic shapes~\cite{Heyde2011,Togashi:2016yzs,Heyde:2016sop,Frauendorf:2017ryj,Zhou:2016ujx}.

\begin{figure}[h!]
\begin{center}
\includegraphics[width=0.9\linewidth]{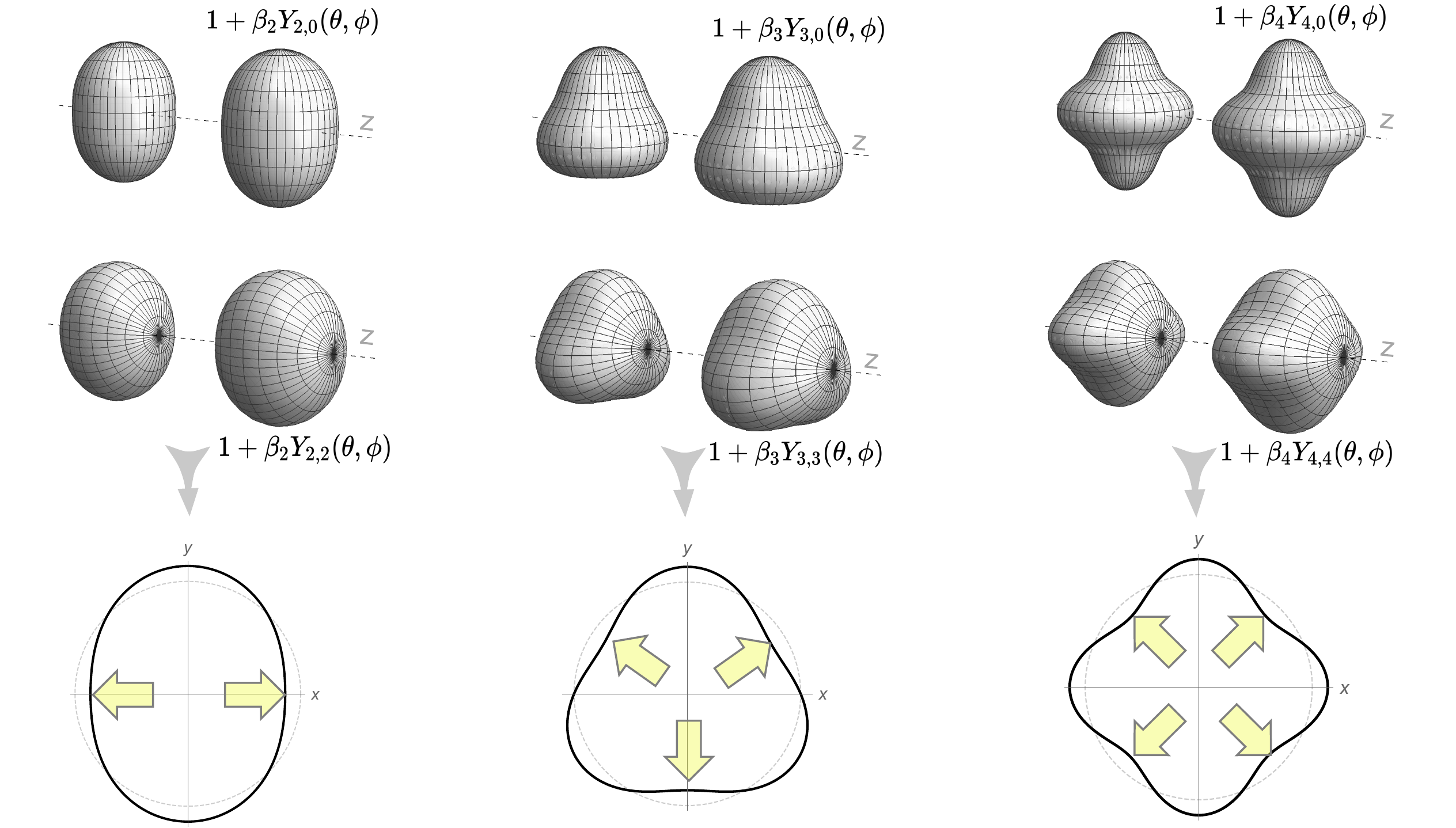}
\end{center}
\caption{\label{fig:idea} The cartoon of collision of nuclei with quadrupole (left), octupole (middle) and hexadecapole (right) deformations including either the $Y_{n,0}$ mode (top row) or the $Y_{n,n}$ mode (middle row) and with $\beta_n=0.25$. The Lorentz contraction in the $z$-direction, by factor of 100 at RHIC and more than a factor of 1000 at the LHC, are not shown. The bottom row shows how the initial condition of the QGP formed after the collision looks like in the transverse plane. The hallow arrows indicate the direction of maximum pressure gradients along which the medium expand with largest velocity, leading to final state harmonic flow $v_n$ with $n$-fold symmetry.}
\end{figure}

No-one has directly observed the deformed nucleus, however. This is because the nucleus is deformed in the so-called intrinsic (body-fixed) frame, and its wave function in the laboratory frame actually does not pick a particular direction. The typical scattering experiments probe the nuclear form factors averaged over all orientations, and the static deformation appears mostly as an increased surface thickness~\cite{DeJager:1987qc}. On the other hand, high-energy heavy ion collisions at RHIC and the LHC, as illustrated in Fig.~\ref{fig:idea}, can image the shape of the nucleus by colliding them together and looking at the collective expansion of the produced system responding to the geometry of the overlap. In these collisions, two Lorentz-contracted nuclei, by a factor of 100 at RHIC and more than a factor of 1000 at the LHC, cross each other over a time scale $\tau<0.1$fm/$c\approx 3\times10^{-24}$s, forming a hot and dense quark-gluon plasma (QGP)~\cite{Busza:2018rrf} in the overlap region, whose initial shape is correlated with the deformed shape of the nuclei. Driven by the large pressure gradient forces, the QGP expands hydrodynamically, converting the spatial anisotropies into azimuthal anisotropies of final-state particles in the momentum space~\cite{Heinz:2013wva}. Nuclear shape imaging is possible because each collision probes simultaneously the entire mass distribution of the nuclei, and one can use particle correlations among thousands of produced particles to infer the two-point and multi-point correlations of this mass distribution and hence its spatial shape. Since the time scales involved in these collisions are much shorter ($<10^{-24}$s) than the typical timescale of the rotational bands ($10^{-21}$s~\cite{Nakatsukasa:2016nyc}), this raises an important question of whether the manifestation of nuclear deformation -- a collective feature of the nuclear many-body system -- is the same across energy scales~\cite{Giacalone:2021udy}.

The shape of the nucleus in nuclear physics is often modeled though a nucleon density profile of the Woods-Saxon form,
\begin{align}\label{eq:1}
\rho(r,\theta,\phi)=\frac{\rho_0}{1+e^{\left[r-R(\theta,\phi)/a_0\right]}},\;R(\theta,\phi) = R_0\left(1+\beta_2 [\cos \gamma Y_{2,0}+ \sin\gamma Y_{2,2}] +\beta_3 \sum_{m=-3}^{3}\alpha_{3,m}Y_{3,m}+\beta_4  \sum_{m=-4}^{4}\alpha_{4,m}Y_{4,m}\right),
\end{align}
where the nuclear surface $R(\theta,\phi)$ is expanded in spherical harmonics $Y_{n,m}$, but keeping only the terms that are most relevant in nuclear structure physics, i.e. quadrupole $n=2$, octupole $n=3$ and hexadecapole $n=4$. Note that the $Y_{2,-1},Y_{2,1}$ and $Y_{2,-2}$ are used to define the intrinsic frame, leaving $Y_{2,0}$ and $Y_{2,2}$ as the only relevant quadrupole components (higher order deformations defined in this frame should have all components as relevant degrees of freedom). The positive number $\beta_2$ describes the overall quadrupole deformation, and the triaxiality parameter $\gamma$ controls the relative order of the three radii $R_a,R_b,R_c$ of the nucleus in the intrinsic frame. It has the range $0\leq\gamma\leq\pi/3$, with $\gamma=0$, $\gamma=\pi/3$, and $\gamma=\pi/6$ corresponding, respectively, to prolate ($R_a=R_b<R_c)$, oblate ($R_a<R_b=R_c$) or maximum triaxiality ($R_a<R_b<R_c$ and $2R_b=R_a+R_c$). Similarly, $\beta_3$ and $\beta_4$ control the overall octupole and hexadecapole deformations, respectively. The $\alpha_{3,m}$ and $\alpha_{4,m}$, in analogue to $\gamma$, are internal ``angular'' parameters describing deviation from axial and/or reflection symmetry, and they satisfy the normalization condition $\sum_{m=-3}^{3} \alpha_{3m}^2=1$ and $\sum_{m=-4}^{4} \alpha_{4m}^2=1$. 

In heavy ion collisions, the initial condition and dynamics of QGP are naturally formulated in a cylindrical coordinate system with the $z$-axis coincide with the beam-line. The initial condition is determined by the distribution of nucleons in the transverse plane $\rho(r_{\perp},\phi)$, which drives the collective flow of final-state particles, reflected by the momentum spectra $N(\pT,\phi)$. The $N(\pT,\phi)$ is often analyzed in terms of a Fourier expansion $dN/d\phi\propto 1+2\sum_nv_n(\pT) \cos\,n(\phi-\Psi_n(\pT))$. The $\rho(r_{\perp},\phi)$ is fully characterized via a 2D multi-pole expansion, whose leading radial modes have the following expression~\cite{Joslin:1983,Teaney:2010vd} in the center of mass frame,
\begin{align}\label{eq:3}
\varepsilon_1e^{i\Phi_1} = -\int d^2r_{\perp} r_{\perp}^3 e^{i\phi}\rho(\vec{r}_{\perp}) / \int d^2r_{\perp}  r_{\perp}^3\rho(\vec{r}_{\perp})\;,\;\varepsilon_ne^{in\Phi_n}|_{n>1} = - \int d^2r_{\perp}  r_{\perp}^n e^{in\phi} \rho(\vec{r}_{\perp}) /  \int d^2r_{\perp}  r_{\perp}^n \rho(\vec{r}_{\perp}),
\end{align}
The 2D eccentricity vectors $\varepsilon_ne^{in\Phi_n}$ are a close analogue of $Y_{n,m}$ in the 3D. In fact, the eccentricity vectors for $n>1$ are directly related to the multi-pole moments of the mass distribution, $\varepsilon_ne^{in\Phi_n} \propto -\lr{Y_{n}^{n}}$, which I will show later lead to a simple relation between $\varepsilon_n$ and $\beta_n$. Note that the radial weight of $\varepsilon_1$ is $r_{\perp}^3$ instead of the naively expected $r_{\perp}$, because the latter contribution vanishes in the center of mass frame and $r_{\perp}^3$ weighting gives the next radial mode.

Study of the relation between the initial-state $\varepsilon_n$ and final-state $v_n$,  within the relativistic viscous hydrodynamics or transport model framework, has always been one central focus of the heavy ion community. Comprehensive model and data comparisons~\cite{Niemi:2015qia} show very good linear relations, $v_n=k_n\varepsilon_n$ not only on average but also for each event~\footnote{The linear relation is very good for $n=2$ and 3 in general and for $n=1$ and $n=4$ in the case of central collisions~\cite{Teaney:2012ke}.}, where the response coefficients $k_n$ capture the transport properties of the QGP produced in the collision. Thanks to the precision measurements of $v_n$ and its event-by-event fluctuations $p(v_n)$~\cite{Jia:2014jca,Busza:2018rrf}, and detailed understanding of the properties of $k_n$~\cite{Teaney:2012ke,Bernhard:2016tnd,Bernhard:2019bmu,Nijs:2020ors}, hydrodynamic models can now determine the $\varepsilon_n$ and $p(\varepsilon_n)$ with enough precision to constrain the deformation parameters. 

Influence of nuclear deformation on dynamics of heavy ion collisions has been considered early on~\cite{Rosenhauer:1986tn,Li:1999bea,Gupta:2000si}. More recent studies focused on the relation between $\beta_2$ and $v_2$~\cite{Heinz:2004ir,Filip:2009zz,Shou:2014eya,Goldschmidt:2015kpa,Giacalone:2017dud,Giacalone:2018apa}. Experimental evidences for quadrupole deformation appear as large differences of $v_2$ between ultra-central collisions (UCC) of different systems, in particular between $^{197}$Au+$^{197}$Au and $^{238}$U+$^{238}$U collisions at RHIC~\cite{Adamczyk:2015obl} and between $^{129}$Xe+$^{129}$Xe and $^{208}$Pb+$^{208}$Pb collisions at the LHC~\cite{Acharya:2018ihu,Sirunyan:2019wqp,Aad:2019xmh}. Ref.~\cite{Giacalone:2021udy} explored the parametric dependence of various flow observables on $\beta_2$, and find that both $\epsilon_2^2$ and $v_2^2$ depend linearly on $\beta_2^2$; a simple formula is derived relating the $\beta_2$ in the two collision systems to the ratio of $v_2$. The influence of octupole deformation is considered recently in Pb+Pb collisions to explain the order of $v_2$ and $v_3$ in the UCC region~\cite{Carzon:2020xwp}.

Another observable showing a strong sensitivity to the nuclear deformation is the Pearson correlation coefficient, $\rho(v_2^2,[\pT])$, between $v_2$ and the mean transverse momentum, $[\pT]$, which probes both the $\beta_2$~\cite{Giacalone:2019pca,Giacalone:2020awm} and its triaxiality $\gamma$~\cite{Jia:2021wbq} of the colliding ions. Recent measurement from the STAR collaboration~\cite{jjia} established unambiguously the large and dominating influence of the nuclear quadrupole deformation of $^{238}$U. The large prolate deformation of $^{238}$U yields a strong negative contribution to the $\rho(v_2^2,[\pT])$, enough to make it change sign. Large influence of deformation is also observed in the fluctuations of $[\pT]$~\cite{jjia}.

Continuing this line of work, I explore the parametric dependence of $\epsilon_1$, $\epsilon_2$, $\epsilon_3$ and $\epsilon_4$ on various deformation parameters $\beta_n$ and deviations from axial and reflection symmetries ($\gamma$ and combinations of $\alpha_{n,m}$ in Eq.~\eqref{eq:1}). In the UCC region, the mean square (ms) of $\epsilon_n$, $\lr{\epsilon_n^2}$, are found to be driven primarily by $\beta_n$ of the same order and the triaxiality parameter $\gamma$ only has very modest impact on $\epsilon_2$. Away from the UCC region, $\lr{\epsilon_n^2}$ are insensitive to $\gamma$, but they receive contributions from $\beta_m$ of a different order $m\neq n$. In other words, I establish the following empirical relation,
\begin{align}\label{eq:4}
\lr{\epsilon_n^2}=a'_n+b'_{n}\beta_n^2+\sum_{m\neq n} b'_{n,m}\beta_m^2\;,\;\;n=1,2,3,4,\;m=2,3,4\; \mathrm{and}\; b'_{1}=0\;,
\end{align}
with significant values of $b'_{n,m}$ observed for $b'_{1,3}$, $b'_{1,4}$ and $b'_{3,4}$. Since $v_n\propto \epsilon_n$, one expects similar parametric dependencies to hold between $\lr{v_n^2}$ and $\beta_m$. This simple scaling relation provides a strong motivation for a collision system scan to map out the shape of atomic nuclei in most interesting region of nuclear chart and compare with the knowledge from nuclear structure physics.

\section{Analytical estimate in ultra-central collisions and Glauber model setup}\label{sec:2a}
To gain some intuitive insight on Eq.~\eqref{eq:4} it is useful to demonstrate its validity using a simpler version of nuclear surface. Here I consider a density distribution of liquid-drop model with a sharp surface: $\rho(r,\theta,\phi)=\rho_0$ when $r<R(\theta,\phi)$ and zero otherwise, and I assume that the energy density distribution is given by the distribution of participating nucleons. I limit the discussion to head-on collisions with nearly maximum overlap i.e. the two nuclei not only have zero impact parameter, but are also required to align in a way to ensure the overlap region contains all the nucleons $\npart=2A$. In reality, the selection of UCC events naturally encompasses a wider range of rotation angles and also a finite range of $\npart$, therefore I also study a second case which requires zero impact parameter but independent rotations for the two nuclei. The details of the calculation can be found in Appendix~\ref{sec:app1}.

 As illustrated in Fig.~\ref{fig:idea}, the maximum $\varepsilon_n$ for $Y_{n,0}$ is reached in a ``body-body'' configuration, when the symmetry-axis of the nuclei is perpendicular to the beam. In the case of $Y_{n,n}$, the maximum $\varepsilon_n$ is reached in a ``tip-tip'' configuration, when the $z$-axis of the nuclei is aligned with the beam. For these two configurations, it is straightforward to calculate eccentricities, they are listed in the first two rows of Table~\ref{tab:2}. However, one is more interested in the eccentricity values averaged over random orientations. The first non-trivial and the most important moment is $\lr{\varepsilon_n^2}$, which relates directly to the $\lr{v_n^2}$ measured by the two-particle correlation method. The results obtained by requiring same random rotations for the two nuclei are listed in the third row of Table~\ref{tab:2}, and those obtained by requiring independent random rotations the two nuclei are listed in the last row of Table~\ref{tab:2}. The two cases have the same dependencies but the coefficients are a factor of two smaller in the second case. 

\begin{table}[!h]
\centering
\small{\begin{tabular}{c|c|c|c|c}
\renewcommand{\arraystretch}{2.2}%
              & $n=1$ & $n=2$ & $n=3$ &$n=4$\\\hline
 & \multirow{2}{*}{$\frac{16}{5\pi\sqrt{7\pi}}\beta_3+$} & \multirow{2}{*}{$\sqrt{\frac{45}{16\pi}}\beta_2+\frac{15}{112\pi}\beta_2^2+$} &\multirow{2}{*}{$\frac{16}{\pi\sqrt{7\pi}}\beta_3+$}& \multirow{2}{*}{$\frac{35}{16\sqrt{\pi}}\beta_4+\frac{45}{16\pi}\beta_2^2+$} \\
$1+\sum_{m=2}^4\beta_mY_{m,0}$   & &&&\\
 & \multirow{2}{*}{\!$\frac{302}{3\sqrt{35}\pi^2}\beta_2\beta_3\!+\!\frac{493967\sqrt{7}}{73920\pi^2}\beta_3\beta_4$\!} & \multirow{2}{*}{${\frac{9\sqrt{5}}{7\pi}\!\beta_2\beta_4+\!\frac{75}{77\pi}\!\beta_4^2\!+\!\frac{1}{\pi}\beta_3^2}$} &\multirow{2}{*}{$\frac{2\sqrt{35}}{3\pi^2}\!\beta_2\beta_3\!+\!\frac{2067\sqrt{7}}{704\pi^2}\!\beta_3\beta_4$}& \multirow{2}{*}{$\frac{215\sqrt{5}}{352\pi}\beta_2\beta_4+\frac{315}{176\pi}\beta_3^2+\frac{43305}{36608\pi}\beta_4^2$} \\
\multirow{2}{*}{$\varepsilon_n$ (body-body)}   & &&&\\
    & $=0.22\beta_3+$&\!$\!=\!0.95\beta_2\!+\!0.043\beta_2^2 +$&\!$=1.08\beta_3+$&$=1.23\beta_4+0.90\beta_2^2+$\\
    & $1.7\beta_2\beta_3+1.8\beta_3\beta_4$&$0.92\beta_2\beta_4+\!0.31\beta_4^2\!+\!0.32\beta_3^2\!$&$0.40\beta_2\beta_3+0.79\beta_3\beta_4$&$0.43\beta_2\beta_4+0.57\beta_3^2+0.38\beta_4^2$\\
 & & & &\\\hline
    &\multirow{3}{*}{$\frac{320}{\sqrt{378}\pi^2}\beta_2\beta_3+\frac{800}{33\pi^2\sqrt{2}}\beta_3\beta_4$}  & \multirow{3}{*}{$\!\sqrt{\frac{15}{4\pi}}\beta_2\!+\!\frac{15}{\sqrt{21}\pi}\beta_2\beta_4\!-\!\frac{15}{8\pi}\beta_2^2\!$} &\multirow{2}{*}{$\frac{64}{\pi\sqrt{70\pi}}\beta_3-$}& \multirow{2}{*}{$\sqrt{\frac{35}{4\pi}}\beta_4+\frac{15}{4\pi}\beta_2^2-$}\\
$1+\sum_{m=2}^4\beta_mY_{m,m}$   & &&&\\
    &\multirow{3}{*}{$=1.67\beta_2\beta_3+1.74\beta_3\beta_4$}  & \multirow{3}{*}{$\!=\!1.09\beta_2+1.0\beta_2\beta_4-0.6\beta_2^2\!$} &\multirow{2}{*}{$\frac{12\sqrt{6}}{\sqrt{7}\pi^2}\beta_2\beta_3-\frac{81}{16\sqrt{2}\pi^2}\beta_3\beta_4$}& \multirow{2}{*}{$\frac{\sqrt{525}}{4\pi}\beta_2\beta_4-\frac{105}{32\pi}\beta_4^2$}\\
\multirow{2}{*}{$\varepsilon_n$ (tip-tip)} & & & &\\
 & $$ &$$ & $\!=1.37\beta_3-\!$&$=1.67\beta_4+1.19\beta_2^2-$\\
 & $$ &$$ & $1.13\beta_2\beta_3-0.36\beta_3\beta_4$ & $-1.82\beta_2\beta_4-1.04\beta_4^2$\\\hline
& & & &\\ $\lr{\varepsilon_n^2}$& \multirow{2}{*}{$\frac{4096}{3675\pi^3}\beta_3^2=0.036\beta_3^2$} & \multirow{2}{*}{$\frac{3}{2\pi}\beta_2^2=0.477\beta_2^2$} & \multirow{2}{*}{$\frac{4096}{245\pi^3}\beta_3^2=0.539\beta_3^2$} &\multirow{2}{*}{ $\frac{35}{18\pi}\beta_4^2+\frac{45}{14\pi^2}\beta_2^4$}\\
 \multirow{2}{*}{(same rotation)}& &&& \\
 &$$ &$$&$$&$=0.62\beta_4^2+0.32\beta_2^4$ \\\hline
& & & &\\ $\lr{\varepsilon_n^2}$& \multirow{2}{*}{$\frac{2048}{3675\pi^3}\beta_3^2=0.018\beta_3^2$} & \multirow{2}{*}{$\frac{3}{4\pi}\beta_2^2=0.239\beta_2^2$} & \multirow{2}{*}{$\frac{2048}{245\pi^3}\beta_3^2=0.270\beta_3^2$} &\multirow{2}{*}{ $\frac{35}{36\pi}\beta_4^2+\frac{45}{28\pi^2}\beta_2^4$}\\
\multirow{2}{*}{ (indep. rotation)}& &&& \\
 &$$ &$$&$$&$=0.31\beta_4^2+0.16\beta_2^4$ \\\hline
\end{tabular}}\normalsize\vspace*{-0.1cm}
\caption{\label{tab:2} The value of eccentricity generated by the deformation component $Y_{n,0}$ (first row) and $Y_{n,n}$ (second row) for the special alignment of two colliding nuclei similar to those shown in Fig.~\ref{fig:idea} which maximizes the eccentricity values, obtained within an optical Glauber model with sharp surface by setting $a_0=0$ in Eq.~\eqref{eq:1}. Here only the leading and subleading order contributions are included. The leading-order mean square values $\lr{\varepsilon_n^2}$ obtained by averaging over common random orientations for the two nuclei and independent random orientations for the two nuclei are given in the third row and the last row, respectively. The values in the latter case are a factor of two smaller, but in both cases they are independent of $\gamma$, $\alpha_{3,m}$ or $\alpha_{4,m}$.}
\end{table}

A few remarks are in order. The maximum possible $\varepsilon_n$ values are different between $Y_{n,n}$ and $Y_{n,0}$, and they are generally comparable to the corresponding $\beta_n$ value. However, the ms values after averaging over random orientations have exactly the same quadratic dependence on $\beta_n$, showing no explicit dependence on the internal angular variables $\gamma$ and $\alpha_{n,m}$ in Eq.~\ref{eq:1} to the leading order~\footnote{The actual probability density distribution $p(\varepsilon_n)$ is different between deformation described by $Y_{n,n}$ and by $Y_{n,0}$. This difference can be captured by the fourth- and higher-order cumulants.}. Remarkably, the octupole deformation also gives rise to a dipolar eccentricity, following the same quadratic dependence on $\beta_3$ but with a coefficient that is a factor of ten smaller. Furthermore, the quadrupole deformation gives rise to a quartic contribution to $\varepsilon_4$, and in an analogy to the well-known non-linear contribution of $\varepsilon_2$ to $\varepsilon_4$~\cite{Teaney:2010vd}, scales as $\lr{\varepsilon_4^2}\approx (0.7-1.4)\lr{\varepsilon_2^2}^2$. Lastly, the coefficients $b_{n,m}'$ listed in the table are derived under a simplified scenario. In a more realistic Monte-Carlo Glauber model calculation based on the Woods-Saxon nuclear profile, the coefficients in the UCC region as shown in bottom row of Fig.~\ref{fig:1}, are comparable or slightly smaller than those obtained by requiring zero impact parameter and independent rotations. 

For a more realistic estimation of influence of nuclear deformation, a Monte-Carlo Glauber model~\cite{Miller:2007ri} is used to simulate collisions of $^{238}$U and $^{96}$Zr systems and calculate $\varepsilon_n$ in each event. These systems are chosen because the experimental collision data exist already. The nucleons are assumed to have a hard-core of 0.4 fm in radii, with a density described by Eq.~\eqref{eq:1}. The nuclear radius $R_0$ and the  surface thickness $a_0$ are chosen to be $R_0=6.81$~fm and $a_0=0.55$~fm for $^{238}$U and $R_0=5.09$~fm and $a_0=0.52$~fm for $^{96}$Zr, respectively. The nucleon-nucleon inelastic cross-section are chosen to be $\sigma_{\mathrm{nn}}=42$~mb at $\sqrtsnn=200$ GeV. In each collision event, nucleons are generated in each nuclei at a random impact parameter from each other. Each nucleus is then rotated by randomly generated three Euler angles before they are set on a straight line trajectory towards each other along the $z$ direction. From this, the nucleons in the overlap region, known as participants, are identified. The $\varepsilon_n$ are calculated from nucleon participants according to Eq.~\eqref{eq:3}, and the results are studied as a function of $\npart$. For a systematic study of the influence of different shapes, one deformation component or particular combination of components of the same $n$ is enabled at a time, the latter is useful to understand the influence of departure from axial and/or reflection symmetry. A special study is performed to also investigate the presence of shapes of different $n$, where two or three non-zero values for $\beta_2$, $\beta_3$ and $\beta_4$ are enabled simultanously.

It is well known that the particle production in nucleus-nucleus collisions only scales approximately with $\npart$. A better scaling can be achieved by considering the constituent quarks as effective degree-of-freedom for particle production~\cite{Adler:2013aqf,Lacey:2016hqy,Loizides:2016djv,Bozek:2016kpf,Acharya:2018hhy}, which would naturally give rise to different $\varepsilon_n$ in each event. Defining centrality with constituent quarks is also expected to change the fluctuations of eccentricity~\cite{Zhou:2018fxx}, and provide a way to quantify the centrality smearing effects. For this purpose, a quark Glauber model from Ref.~\cite{Loizides:2016djv} is used. Three quark constituents are generated for each nucleon according to the ``mod'' configuration~\cite{Mitchell:2016jio}, which ensures that the radial distribution of the three constituents after re-centering follows the proton form factor $\rho_{\mathrm{proton}}(r) = e^{-r/r_0}$ with $r_0=0.234$~fm~\cite{DeForest:1966ycn}. The value of quark-quark cross-section is chosen to be $\sigma_{\mathrm{qq}}=8.2$~mb in order to match the $\sigma_{\mathrm{nn}}$. The  $\varepsilon_n$ are then calculated from the list of quark participants in the overlap region, and the number of quark participants $\nqp$ is used as an alternative centrality estimator. In the quark Glauber model, I also keep track explicitly the participant nucleons, i.e. a nucleon is counted as participant as long as one of its quark participate in the collision.  This paper presents and compares results obtained from both nucleon participants and quark participants. 

In the presence of large deformation, the total volume of the nucleus increases slightly for fixed $R_0$~\cite{Myers:1983seb}. Considering the quadrupole deformation only, for the largest value considered, $\beta_2=0.34$, the ratio to the original volume is approximately $1+\frac{3}{4\pi}\beta_2^2+\frac{\sqrt{5}}{28\pi^{3/2}}\cos(3\gamma)\beta_2^3=1.021+0.0004\cos(3\gamma)$. In order to keep the overall volume fixed, it would require a small less than 1\% decrease of the $R_0$, which is safely ignored in the present study.
\section{Results}\label{sec:4}
The goal of this paper is to explore the relation between $\lr{\varepsilon_n^2}$ and various deformation parameters in Eq.~\eqref{eq:1}, and to provide insights on the deformation dependence of experimentally measured $\lr{v_n^2}$. The influence of nuclear deformation on higher-order cumulants of $\varepsilon_n$ will be explored in a separate study. Section~\ref{sec:41} establishes the quadratic relation Eq.~\eqref{eq:4} by considering the axial-symmetric deformation $Y_{n,0}$, $n=2,3$ and 4. The influences of non-axial deformation, $Y_{n,m\neq0}$, characterized by the triaxiality parameter $\gamma$ and $\alpha_{n,m}$ parameters are considered in Section~\ref{sec:42}. One finds that the slope parameters $b_{n,m}'$ have a very weak dependence on the $\gamma$ and $\alpha_{n,m}$. Section~\ref{sec:43} considers the presence of multiple shape components $\beta_2$, $\beta_3$, and $\beta_4$, which is generally expected in nuclear structure physics. For moderate deformation values, one finds that the non-linear contributions, terms like $\beta_n\beta_m, m\neq n$ is subdominant and Eq.~\eqref{eq:4} still holds well. Section~\ref{sec:44} discusses ways to constrain these deformations simultaneously using flow measurements.
 
\subsection{Influence of axial-symmetric multipole deformation}\label{sec:41}
\begin{figure}[h!]
\begin{center}
\includegraphics[width=0.9\linewidth]{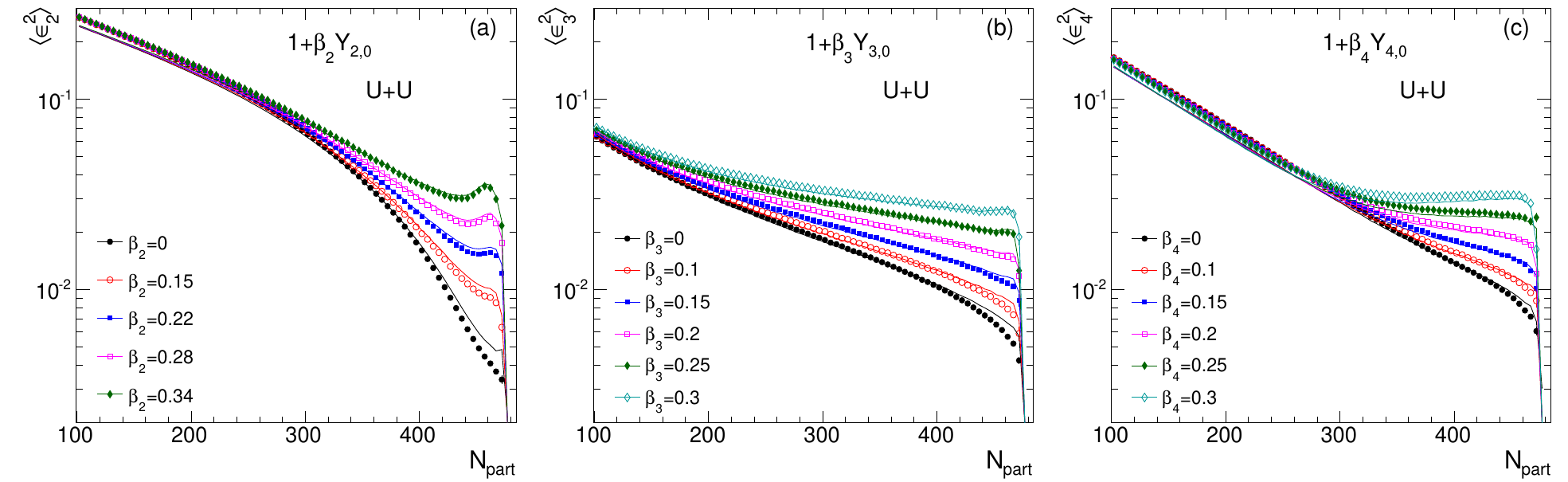}
\includegraphics[width=0.9\linewidth]{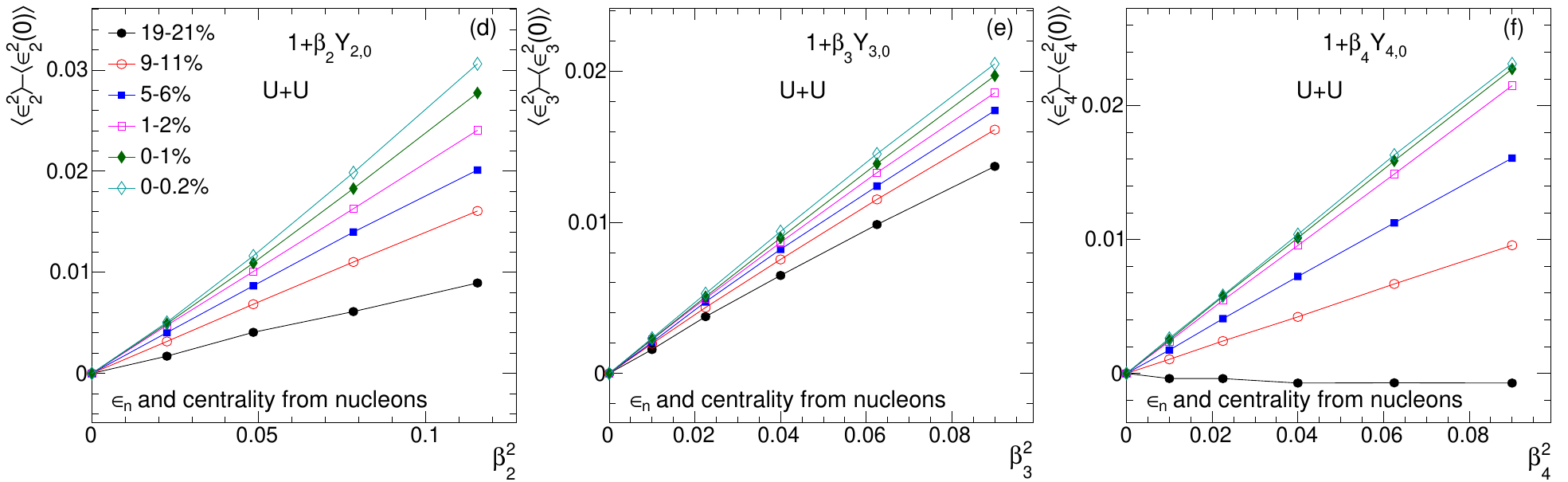}
\includegraphics[width=0.9\linewidth]{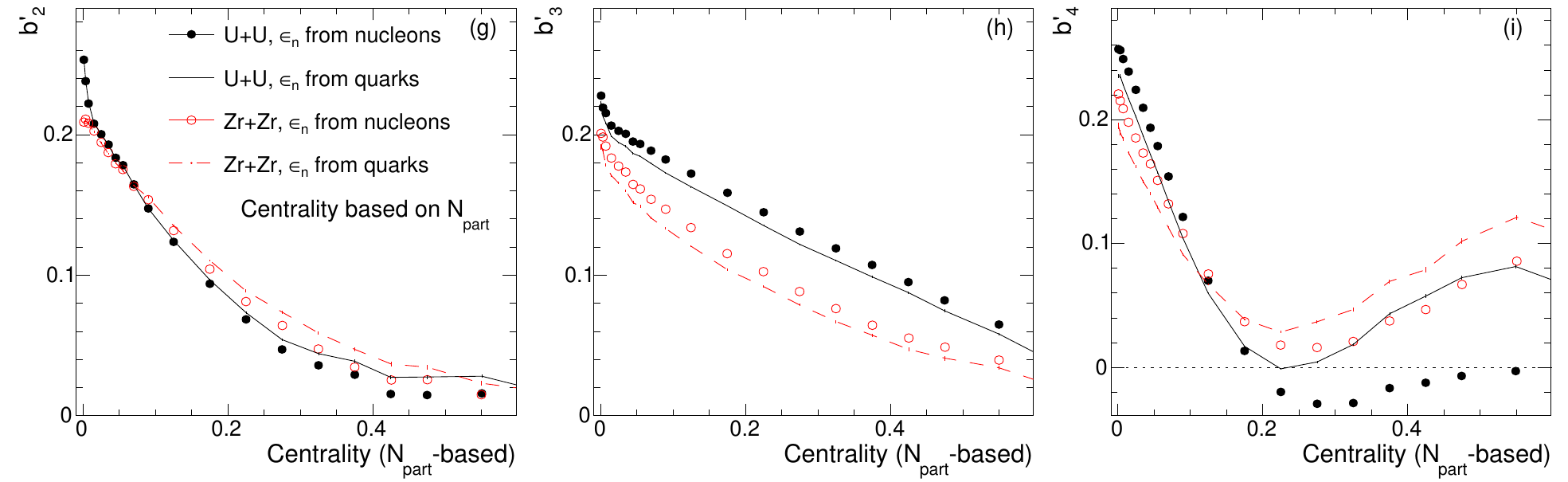}
\end{center}
\caption{\label{fig:1} Top row: The $\npart$ dependence of mean squre eccentricity $\lr{\varepsilon_n^2}$ for several $\beta_n$ values considering only the $Y_{n,0}$ component, with markers and lines correspond $d_{\perp}$ obtained with nucleons and quarks, respectively. Middle row: The $\beta_n^2$ dependence of $\lr{\varepsilon_n^2}$ in several centrality ranges based on $\npart$, which can be nicely described by a linear function $\lr{\varepsilon_n^2}-\lr{\varepsilon_n^2}_{\beta_n=0} = b'_n \beta_n^2$. Bottom row: the centrality dependence of the slope parameter $b'_n$ in U+U (black) and Zr+Zr (red) systems for $\varepsilon_n$ calculated based on nucleons (markers) or quarks (lines). The results are shown for $n=2$ (left column), $n=3$ (middle column) and $n=4$ (right column). The first three points in the bottom panels correspond to 0--0.2\%, 0.2--0.5\% and 0.5--1\%, respectively.}
\end{figure}

The top row of Fig.~\ref{fig:1} shows the $\npart$ dependence of $\lr{\varepsilon_n^2}$, $n=2$, 3 and 4 for various $\beta_n$ values in U+U collisions, calculated from the participating nucleons according to Eq.~\eqref{eq:3}. One observes that the larger $\beta_2$ values increase $\varepsilon_2$ mostly in the central region, while larger $\beta_3$ values increase $\varepsilon_3$ over the full centrality range. The reason is that the contribution associated with the average elliptic geometry to $\varepsilon_2$ dominates over the deformation effects in the mid-central and peripheral collisions. On the other hand, the $\varepsilon_3$ without nuclear deformation arises solely from random fluctuations of nucleon positions and has a much smaller value, therefore the $\varepsilon_3$ is more sensitive to $\beta_3$. The increase of $\varepsilon_4$ with $\beta_4$ is observed only in the central region. In fact, the peripheral region shows a slight decrease of $\varepsilon_4$ with $\beta_4$. Overall, the influence of deformation on $\varepsilon_n$ is largest in the most central region for all harmonics.

In the same plots, I also show the $\lr{\varepsilon_n^2}$ calculated from quark participants as solid lines, with the same color as those calculated from nucleon participants. Small differences are observed in the UCC region when $\beta_n$ are small, or in the peripheral region for $n=2$, implying that the influences of deformation are insensitive to nucleon substructures.   

To quantify these dependencies, $\lr{\varepsilon_n^2}$ values obtained for fixed $\npart$ are averaged in narrow centrality ranges, and plotted as a function of $\beta_n^2$ in the middle row of Fig.~\ref{fig:1}. An linear dependence is observed in all cases, confirming the first part of the relation in Eq.~\eqref{eq:4} involving $a'_n$ and $b'_n$. Note that $a'_n$ correspond to eccentricities in the absence of deformation $a'_n=\lr{\varepsilon_n^2}_{|\beta_n=0}$ shown by the black solid circles in the top row, while $b'_n$ describes the slope of the $\beta_n^2$ dependencies in the middle row .

The bottom row of Fig.~\ref{fig:1} shows the centrality dependence of $b'_n$ for U+U and Zr+Zr collisions. The values of $b'_n$ are largest in the UCC region and decrease toward mid-central and peripheral region. It is quite remarkable that the value of $b'_n$ starts at around 0.2--0.3 for all harmonics in both collision systems. This value of $b'_n$ reflects an effect that is purely geometrical. If the two nuclei were to collide head-on in the direction perpendicular to maximum deformation as shown in Fig.~\ref{fig:idea}, $b'_n$ should be on the order of unity, see Table~\ref{tab:2}. In reality, after averaging over all possible random orientations, the $b'_n$ values are reduced to about 0.2--0.3. In a Monte-Carlo Glauber with finite number of nucleons, the random fluctuation of nucleon positions smear the correlation between the shape of the overlap region and the $\npart$. This smearing is expected to be larger for smaller system, leading to a slightly smaller $b'_n$ in the Zr+Zr collisions than in the U+U collisions. Note that the centrality and collision system dependencies of $b_n'$ are just the opposite of $a'_n$ (the latter corresponds to the $\lr{\varepsilon_n^2}$ without deformation shown in the top row). The values for $a'_n$ are smallest in the UCC region and increase towards more peripheral region and exhibit a much larger difference between Zr+Zr and U+U.

\begin{figure}[h!]
\begin{center}
\includegraphics[width=0.9\linewidth]{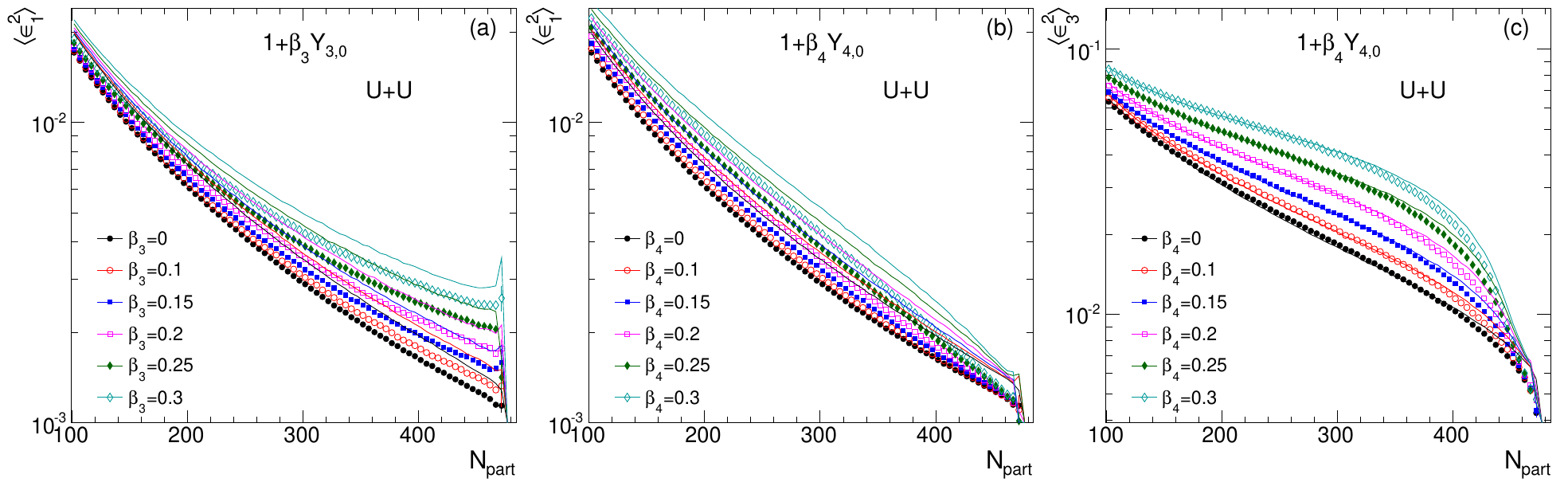}
\includegraphics[width=0.9\linewidth]{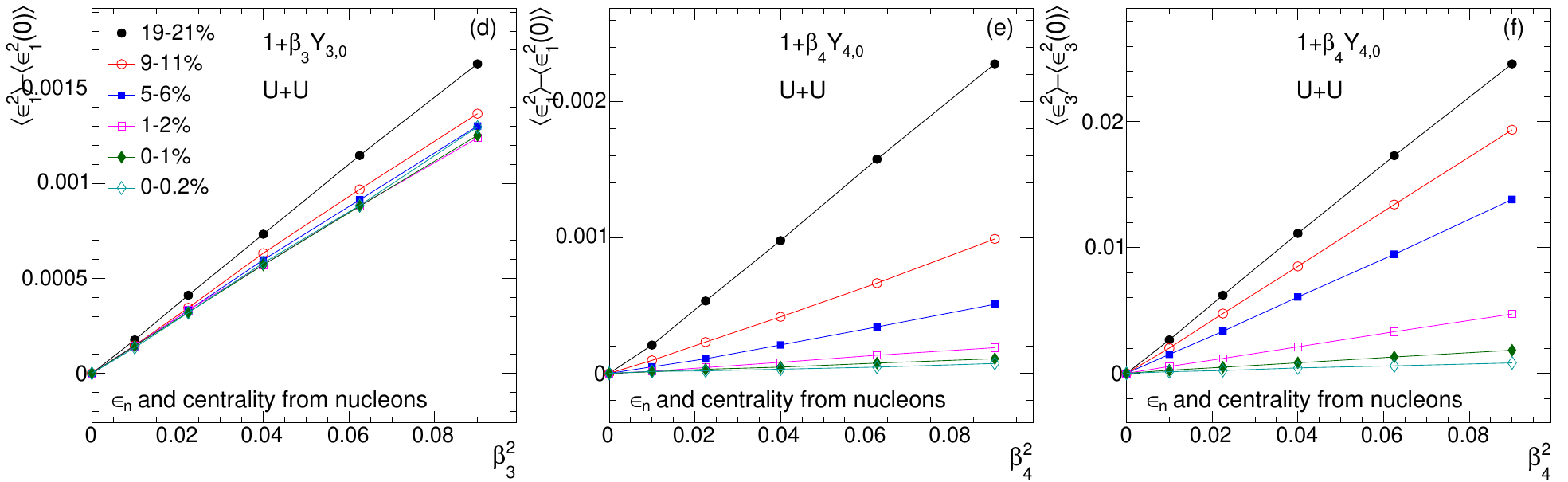}
\includegraphics[width=0.9\linewidth]{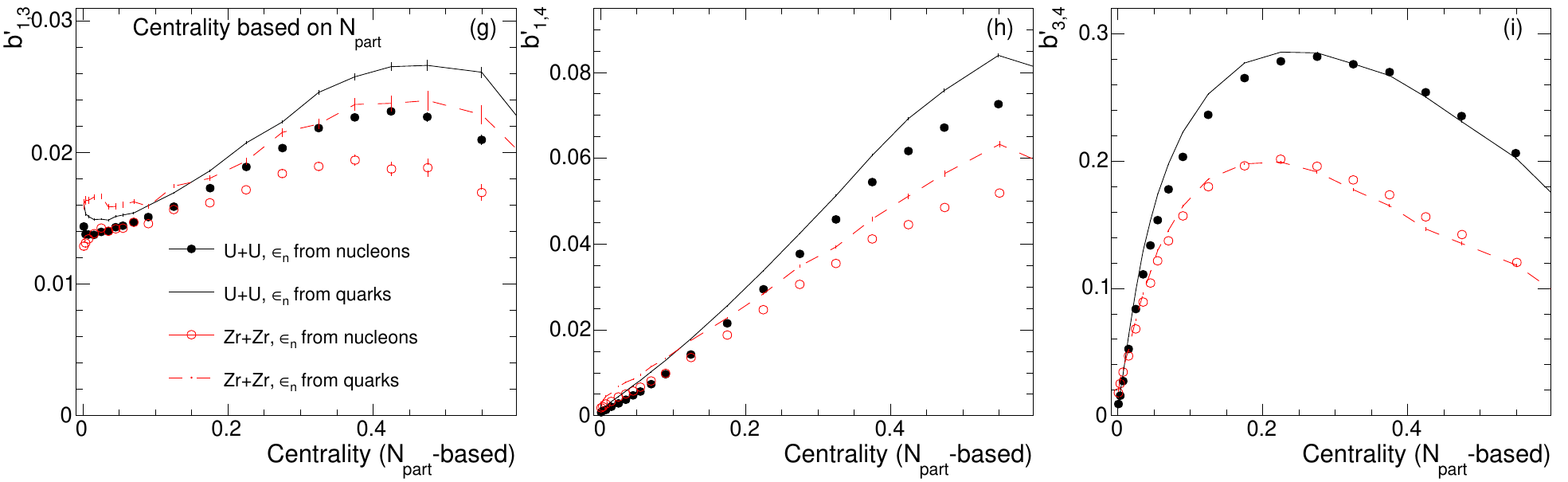}
\end{center}
\caption{\label{fig:2} Characterization of the influence of $\beta_3$ to $\lr{\varepsilon_1^2}$ (left column),  $\beta_4$ to $\lr{\varepsilon_1^2}$ (middle column) and $\beta_4$ to $\lr{\varepsilon_3^2}$ (right column). The top row shows the centrality dependence of $\lr{\varepsilon_n^2}$ for several values of $\beta_3$ or $\beta_4$ as indicated in the panels, with markers and lines correspond $d_{\perp}$ obtained with nucleons and quarks, respectively. The middle row shows the $\beta_3^2$ or $\beta_4^2$ dependence of $\lr{\varepsilon_1^2}$ or $\lr{\varepsilon_1^2}$ in several centrality ranges. The bottom row shows the centrality dependence of the extracted slope parameter $b'_{1,3}$ (left), $b'_{1,4}$ (middle) or $b'_{3,4}$ (right) for U+U (black) and Zr+Zr (red) systems for $\varepsilon_n$ calculated based on nucleons (markers) or quarks (lines). The first three points in the bottom panels correspond to 0--0.2\%, 0.2--0.5\% and 0.5--1\%, respectively.}
\end{figure}

The bottom row of Fig.~\ref{fig:1} also compares the $b_n'$ calculated from nucleon participants with that calculated from quark participants. In the central and mid-central collisions, the differences are negligible for $n=2$, but for $n=3$ and $n=4$ the results based on quark participants are systematically smaller. At this point, one may wonder if the $\varepsilon_n$ are also affected by the $\beta_m$ of different order, $m\neq n$. I have performed such calculations.  In most cases, the influences are small. But one identifies three cases for which the influences are quite large. In particular, one finds that the octupole deformation contributes strongly to the dipolar eccentricity in all centrality, and the hexadecapole deformation contributes to both dipolar and triangular eccentricities in the mid-central collisions. They are presented in Fig.~\ref{fig:2} with a similar layout as Fig.~\ref{fig:1}.  As these contributions are a global geometry effect with a radial distribution different from contributions arising from random fluctuation in nucleon positions in each event, they probably will be damped differently by viscous effects in comparison to $a_n'$. Similar effects are known in the context of hydrodynamic model studies as leading and sub-leading eccentricities, which characterize different length scale in radial direction for $\varepsilon_n$, and sub-leading $\varepsilon_n$ with higher frequency in radial direction is more damped than the leading $\varepsilon_n$~\cite{Mazeliauskas:2015vea,Mazeliauskas:2015efa}. Results for other $b_{n,m}'$ can be found in Figs.~\ref{fig:app0}--\ref{fig:app3} in the Appendix~\ref{sec:app2}, including the quartic dependence of $\lr{\varepsilon_4^2}$ on $\beta_2$ predicted in Tab.~\ref{tab:2}.

The middle row of Fig.~\ref{fig:2} shows that contributions between different orders also follow a quadratic dependence, confirming the second part of the Eq.~\eqref{eq:4}. The slopes, $b'_{1,3}$, $b'_{1,4}$ and $b'_{2,4}$, are summarized in the bottom row of Fig.~\ref{fig:2}. One should not be tricked by the apparent small value of $b_{1,3}'$, though. Since the value of $a'_1$, the $\lr{\varepsilon_1^2}$ without deformation, is very small in the UCC region, even a value of $b'_{1,3}=0.015$ together with a modest octupole deformation of $\beta_3=0.1$ could increase the $\lr{\varepsilon_1^2}$ by about 15\%. In mid-central collisions, due to a much larger $a'_1$, the combined contributions from $\beta_3=0.1$ and $\beta_4=0.1$ are less than 10\%. This result suggests that the dipolar flow in the UCC region could in principle be used to probe the octupole deformation. On the other hand, the influence of $\beta_4$ on $\varepsilon_3$ is significant in the mid-central collisions, and is negligible in the UCC region. The bottom row of Fig.~\ref{fig:2} also compares the slope parameters between U+U and Zr+Zr collisions, they are very similar in the UCC region, but values in Zr+Zr are about 20\% smaller in the mid-central collisions.

To summarize the main message of Fig.~\ref{fig:2}, the $\varepsilon_n$ for $n=2$,3, and 4 in the UCC region is not affected by deformation of different order $\beta_m$, $m\neq n$, leading to a particularly simple expression, $\lr{\epsilon_n^2}_{\mathrm{UCC}} = a'_n+b'_{n}\beta_n^2$. Exploiting this relation in the UCC collisions from experimentally measured $v_n$ values provides a clean way to constrain the $\beta_n$ parameters as will be discussed in Section~\ref{sec:44}.

\subsection{Influence of non-axial deformations}\label{sec:42}
\begin{figure}[h!]
\begin{center}
\includegraphics[width=0.8\linewidth]{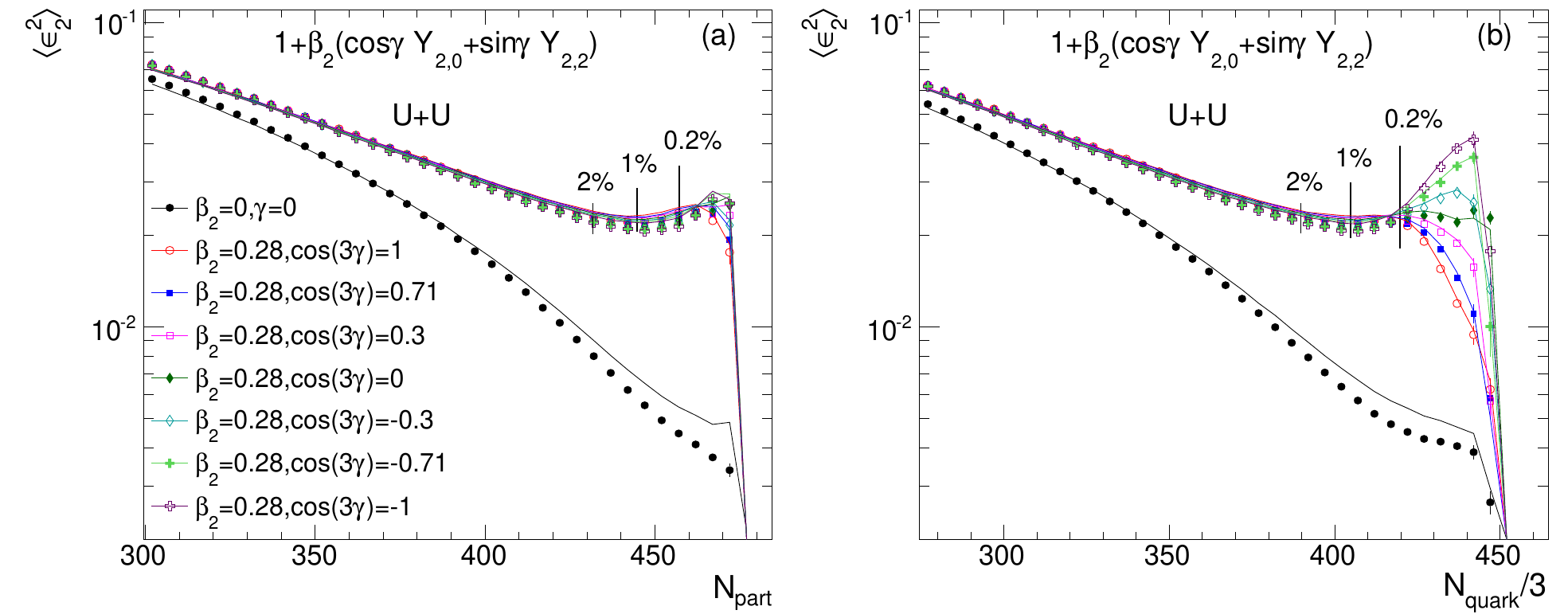}
\includegraphics[width=0.8\linewidth]{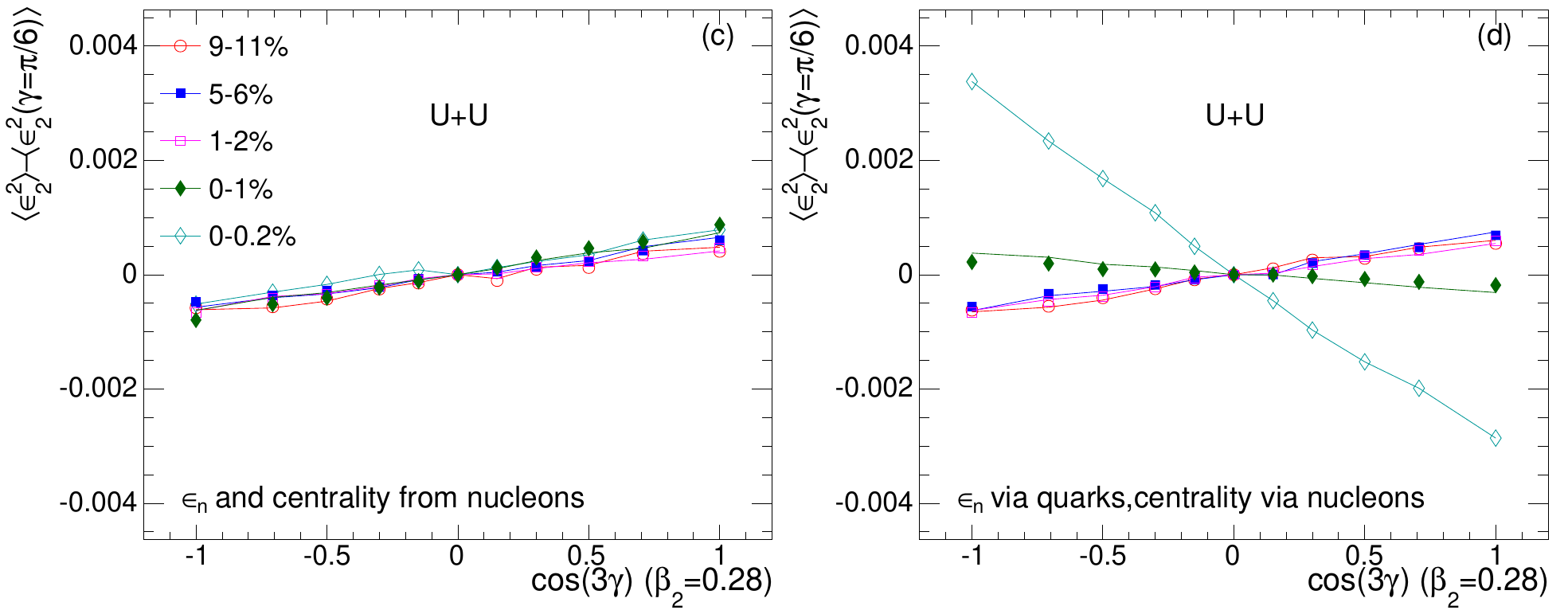}
\includegraphics[width=0.8\linewidth]{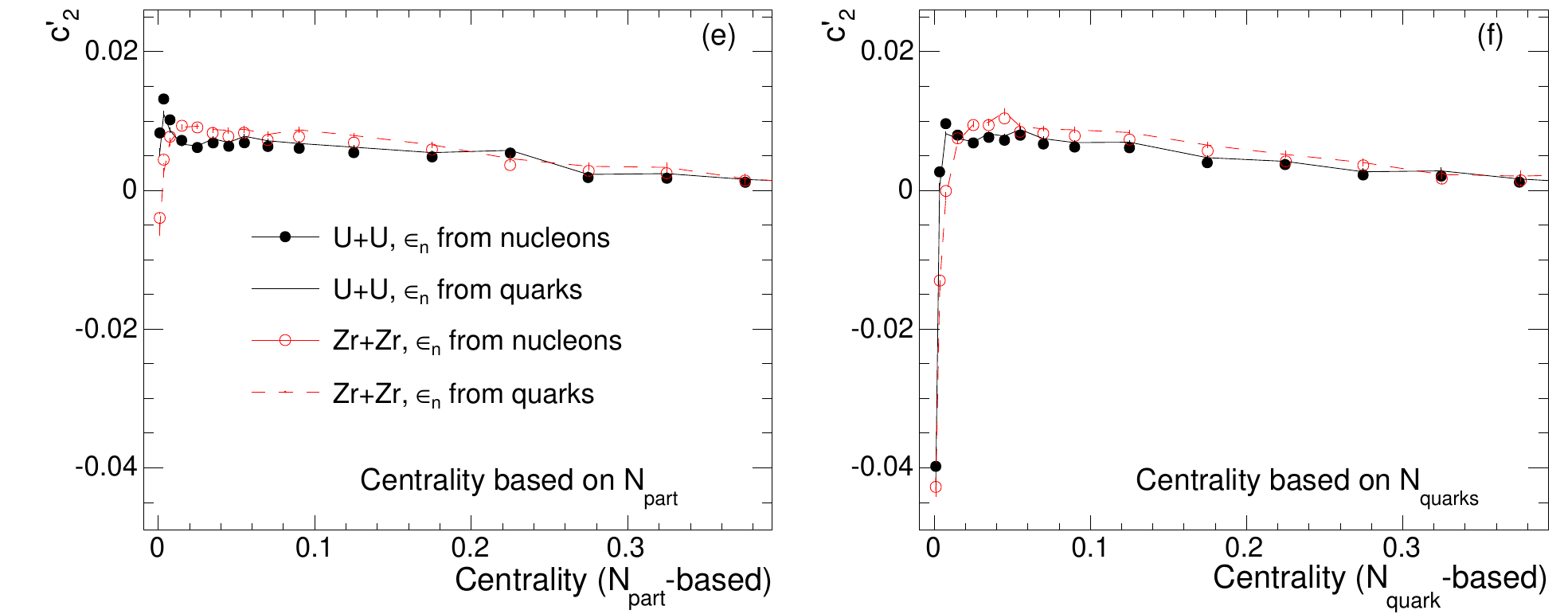}
\end{center}
\caption{\label{fig:3}   Characterization of the influence of triaxiality $\gamma$ on $\lr{\varepsilon_2^2}$. Top row: The $\npart$ (left) and $\nqp$ (right) dependencies of $\lr{\varepsilon_2^2}$ for different values of $\gamma$, where markers and lines correspond $d_{\perp}$ obtained with nucleons and quarks, respectively. Middle row: The  $\lr{\varepsilon_2^2}- \lr{\varepsilon_2^2(\gamma=\pi/6)}$ as a function of $\cos(3\gamma)$ in several centrality ranges based on $\npart$ (left) and $\nqp$ (right), which follows a linear function of $\cos(3\gamma)$, whose slope is parametrized as $c_2'\beta_2^2$ (see Eq.~\eqref{eq:6}). Bottom row: the extracted $c'_2$ as a function of centrality for U+U (black) and Zr+Zr (red) systems for $\varepsilon_n$ calculated based on nucleons (markers) or quarks (lines).  The first three points in the bottom panels correspond to 0--0.2\%, 0.2--0.5\% and 0.5--1\%, respectively.}
\end{figure}

Let's first consider the influence of triaxiality parameter $\gamma$, which mixes the contribution from $Y_{20}$ and $Y_{22}$ components, while keeping the overall magnitude of quadrupole deformation $\beta_2$ fixed. The top-left panel of Fig.~\ref{fig:3} shows the $\npart$ dependence of $\lr{\varepsilon_2^2}$ for $\beta_2=0.28$ but different $\gamma$ values in the U+U collisions. They are contrasted to the case for spherical nuclei $\beta_2=0$. It is clear that over most of the centrality range, $\lr{\varepsilon_2^2}$ have very little sensitivity to $\gamma$. The $\lr{\varepsilon_2^2}$ calculated with quark participants, shown as solid lines in the same panel, also give very similar results.

To test the influence of volume/centrality fluctuations, results obtained using $\nqp$ as centrality is shown in the top-right panel. Large splittings between different $\gamma$ cases are observed in the UCC region of 0--1\%. Namely, the $\lr{\varepsilon_2^2}$ for oblate deformation $\gamma=\pi/3$ shows a stronger increase as a function of $\nqp$ before they all start to decrease slightly at the largest $\nqp$ values. This behavior suggests that the events selection based on $\npart$ or $\nqp$ have different correlation with, and therefore different sensitivity to, the triaxiality of the nucleus. The largest difference is reached between the prolate deformation and the oblate deformation, consistent with a previous study based on the AMPT model~\cite{Giacalone:2021udy,Jia:2021wbq}. In that study, a similar dependence on $\gamma$ is observed for the final-state $v_2$. Interestingly, the U+U $v_2$ data from the STAR Collaboration show a decreasing behavior as a function of $\nch$ in the UCC region, while the Au+Au $v_2$ data show a slight rising trend~\cite{Adamczyk:2015obl}, compatible with a prolate deformation of $^{238}$U nucleus and a oblate deformation of $^{197}$Au. In summary, our results suggest that the two-particle correlators $\lr{v_2^2}$ are sensitive to the triaxiality only in the UCC region, and the level of sensitivity depends on the choice of centrality estimator.

The middle row of Fig.~\ref{fig:3} quantifies the $\gamma$ dependence of $\lr{\varepsilon_2^2}$ in several centrality ranges in $\npart$ on the left and $\nqp$ on the right in the U+U collisions. The $\gamma$ dependence is well described by a linear function of $\cos(3\gamma)$, reflecting the expected three-fold periodicity. Similar observation is also made in the Zr+Zr collisions, although the sensitivity to $\gamma$ is observed over a larger centrality range. Based on this finding, I arrive the following empirical formula that accounts for the dependence on both $\beta_2$ and $\gamma$,
\begin{align}\label{eq:6}
\lr{\varepsilon_2^2} = a'_2+(b'_2+c'_2 \cos(3\gamma)) \beta_2^2\;.
\end{align}
Note that the $c'_2\beta_2^2$ is the slope of the $\cos(3\gamma)$ dependence in the middle row of Fig.~\ref{fig:3}. The bottom row of Fig.~\ref{fig:3} shows the centrality dependence of $c'_2$. The value of $c'_2$ is generally much smaller than $b'_2$, $c'_2\ll b'_2$, and approaches zero in the peripheral collisions. However, its value in the UCC region changes sign and could gain a sizable magnitude depending on the centrality estimator. Lastly, for other eccentricities $\varepsilon_n, n\neq2$, only very small dependencies on the triaxiality are observed, typically less than 5\%. However, these dependencies to a good extent can also be described by a $\cos(3\gamma)$ function (see the left column of Figs.~\ref{fig:app0}-\ref{fig:app3}).

This result begs the question of whether the finding about triaxiality also applies for the octupole and hexadecapole deformations. These higher-order deformations have many more shape parameters. In the intrinsic frame defined by the quadrupole deformation, after taking out the $\beta_n$ that describe the overall strength of the deformation, there are still six and eight independent shape variables for octupole and hexadecapole deformations, respectively. Besides, there are also large redundancies in the parameter space due to spatial symmetry of spherical harmonics. For example, if the underlying quadrupole deformation is axial-symmetric, terms like $\cos(\delta)Y_{n,m}+\sin(\delta)Y_{n,-m}$ can be absorbed by an azimuthal rotation of $Y_{n,m}$ without real physical consequence. There have been several attempts to find efficient parameterizations to reduce this redundancy, see Refs.\cite{Hamamoto:1991vdk,Rohozinski:1997zz}. Our paper follows a more relaxed approach, where I just test special cases of the octupole and hexadecapole shapes. It is reassuring that $\lr{\varepsilon_n^2}$ has very small sensitivity to these internal angular parameters, as will be described below.

\begin{figure}[h!]
\begin{center}
\includegraphics[width=0.8\linewidth]{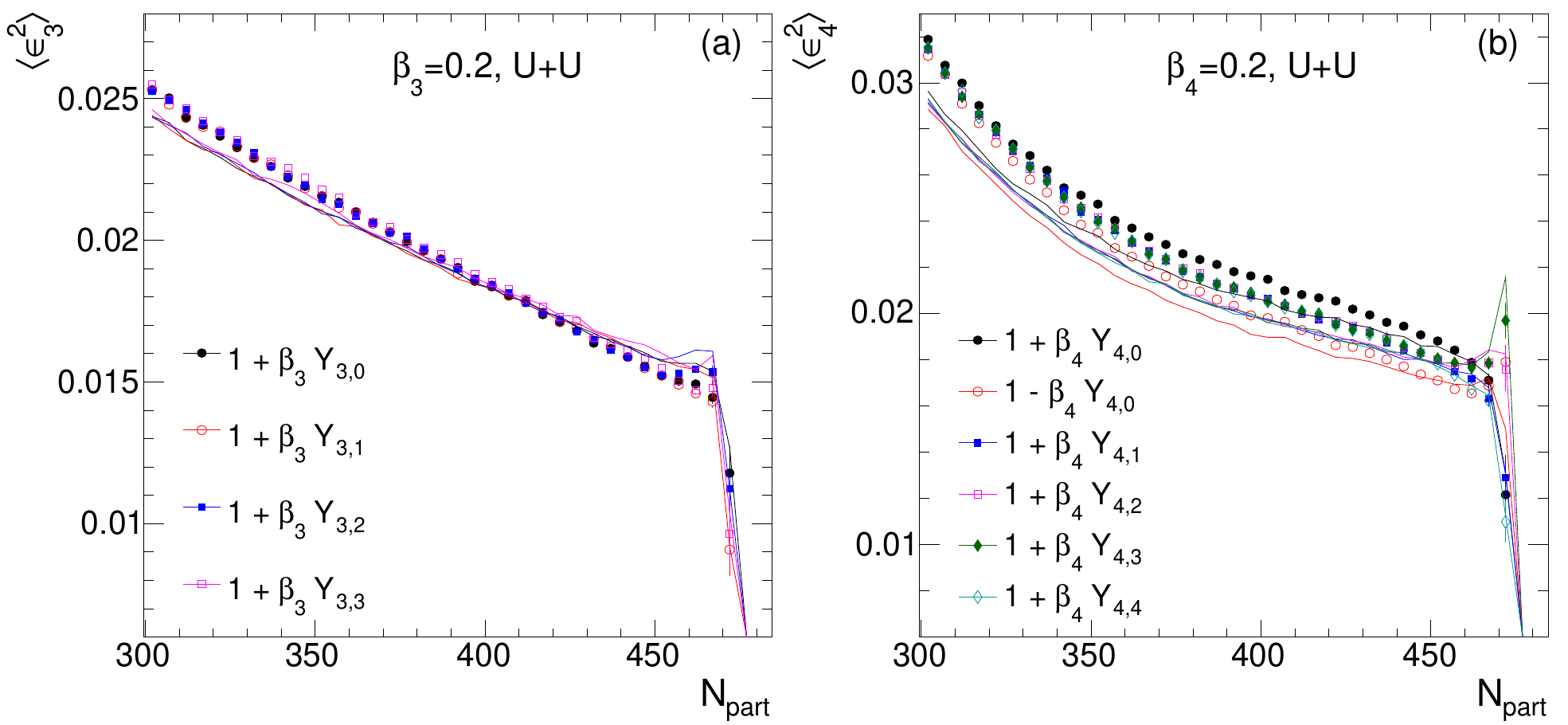}
\end{center}
\caption{\label{fig:4} Left: The $\npart$ dependence of $\lr{\varepsilon_3^2}$ for different component of octupole deformation $Y_{3,m}$ with a deformation value of $\beta_3=0.2$. Right:  The $\npart$ dependence of $\lr{\varepsilon_4^2}$ for different component of octupole deformation $Y_{4,m}$ with a deformation value of $\beta_4=0.2$. The $\varepsilon_n$ are calculated either from nucleons (symbols) or quarks (lines). For all cases $\beta_2=0$.}
\end{figure}

For this study, I consider all real valued spherical harmonics $Y_{3,m}$ and $Y_{4,m}$. They are introduced one at a time in the Glauber model and the resulting $\varepsilon_n$ are calculated. The results are summarized in Fig.~\ref{fig:4}. I found that $\varepsilon_3$ values are the same for all $Y_{3,m}$ components, except for small differences in the UCC region. I also tried several combinations, such as $\cos(\delta)Y_{3,0}+\sin(\delta)Y_{3,1}$ with $\delta$ a free mixing angle, and the conclusion remains the same. I suspect that this is true for general mixing of all components $\alpha_{3,m}$ in Eq.~\eqref{eq:1}, as long as $\sum_{m=-3}^{3} \alpha_{3m}^2=1$ is satisfied. This independence is also observed for $\beta_3$ contribution to the dipolar eccentricity $\varepsilon_1$ and probably is a property for all odd-order deformations. 

On the other hand, the $\lr{\varepsilon_4^2}$ values show a modest, at a level of 15\%, differences among different $Y_{4,m}$ components for $\beta_4=0.2$. Results for non-axial components $1+\beta_4Y_{4,m}, m\neq0$ lie exactly between $1+\beta_4Y_{4,0}$ and $1-\beta_4Y_{4,0}$. The differences are largest in central collisions but are present throughout the entire centrality range. This is different from $\lr{\varepsilon_2^2}$, for which the dependence on the triaxiality is observed only in the UCC region. I initially thought that the $\lr{\varepsilon_4^2}$ for the most general hexadecapole shape should be in between the results for $Y_{40}$ and $-Y_{40}$, which turns out is not the case. In fact, the extrema of $\lr{\varepsilon_4^2}$ are reached for deformation described by $\sqrt{7/12}Y_{4,0}+\sqrt{5/12}Y_{4,4}$ and $\sqrt{5/12}Y_{4,0}-\sqrt{7/12}Y_{4,4}$, with the maximum for $\beta_4=|\beta_4|$ and minimum for $\beta_4=-|\beta_4|$~\cite{Rohozinski:1997zz}. Identifying heavy ion observables that are sensitive to the sign of $\beta_{4}$ will be particularly useful for understanding the nuclear fission data~\cite{{Lemmon:1993nvm}}. A more detailed investigation of this topic is given in the Appendix~\ref{sec:app2}. 

\subsection{Simultaneous presence of quadrupole-octupole-hexadecapole deformations}\label{sec:43}
Although the axial quadrupole distortion is the nuclear deformation of primary importance, secondary contributions from octupole and hexadecapole components often coexist and can be important in some regions of nuclear chart~\cite{Butler:2016rmu}. One example is the pear-shaped $^{224}$Ra~\cite{Gaffney:2013} with a $(\beta_2,\beta_3,\beta_4)$ value of $(0.1545, 0.097, 0.080)$~\cite{Nazarewicz:1984mkb}. A summary of the deformation parameters for the large systems collided at RHIC and the LHC are listed in Tab.~\ref{tab:3}, highlighting the importance of possible higher-order deformations. It would be interesting to study how the eccentricities depend on the simultaneous presence of these different deformations, in particular, whether the contribution from each component to $\varepsilon_n$ is independent of each other.

\begin{table}[!h]
\centering
\begin{tabular}{|c|ccc|}\hline
&$\beta_2$&$\beta_3$&$\beta_4$\\\hline
$^{238}$U& 0.286~\cite{Raman:1201zz}& 0.078~\cite{Agbemava:2016mvz}& 0.07--0.09~\cite{Libert:1982zz,Moller:2015fba}  \\\hline
$^{208}$Pb & 0.05~\cite{Raman:1201zz}&0.04\cite{Robledo:2011nf}&?   \\\hline
$^{197}$Au & -(0.13-0.16)~\cite{Moller:2015fba,Hilaire:2007tmk}&?&-0.03~\cite{Moller:2015fba} \\\hline
$^{129}$Xe& 0.16~\cite{Moller:2015fba}& ? & ?  \\\hline          
$^{96}$Ru  &0.05-0.16~\cite{Moller:2015fba,Raman:1201zz}&?&?  \\\hline
$^{96}$Zr &0.08~\cite{Raman:1201zz}&0.20-0.27~\cite{Kibedi:2002ube}&0.06~\cite{Moller:2015fba} \\\hline
\end{tabular}
\caption{\label{tab:3} Some estimates of the deformation values $\beta_2, \beta_3$, and $\beta_4$ for the large nuclei collided at RHIC and the LHC with references given, mostly based on global analysis of the $B(En)$ transition data.} 
\end{table}

\begin{figure}[h!]
\begin{center}
\includegraphics[width=1\linewidth]{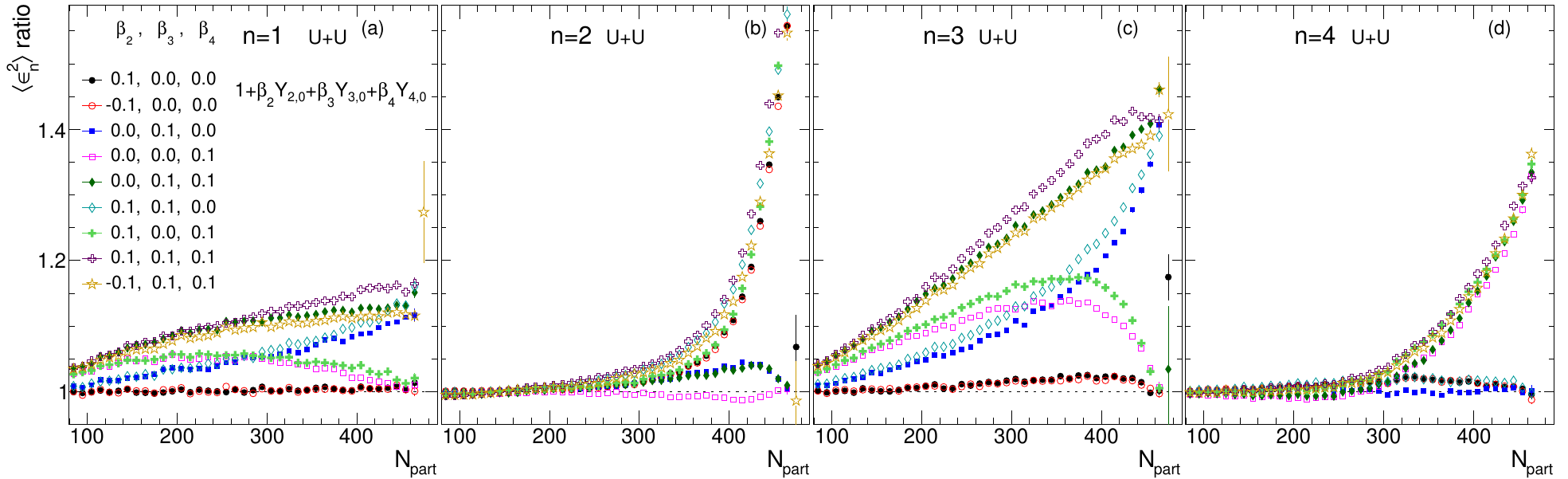}
\includegraphics[width=1\linewidth]{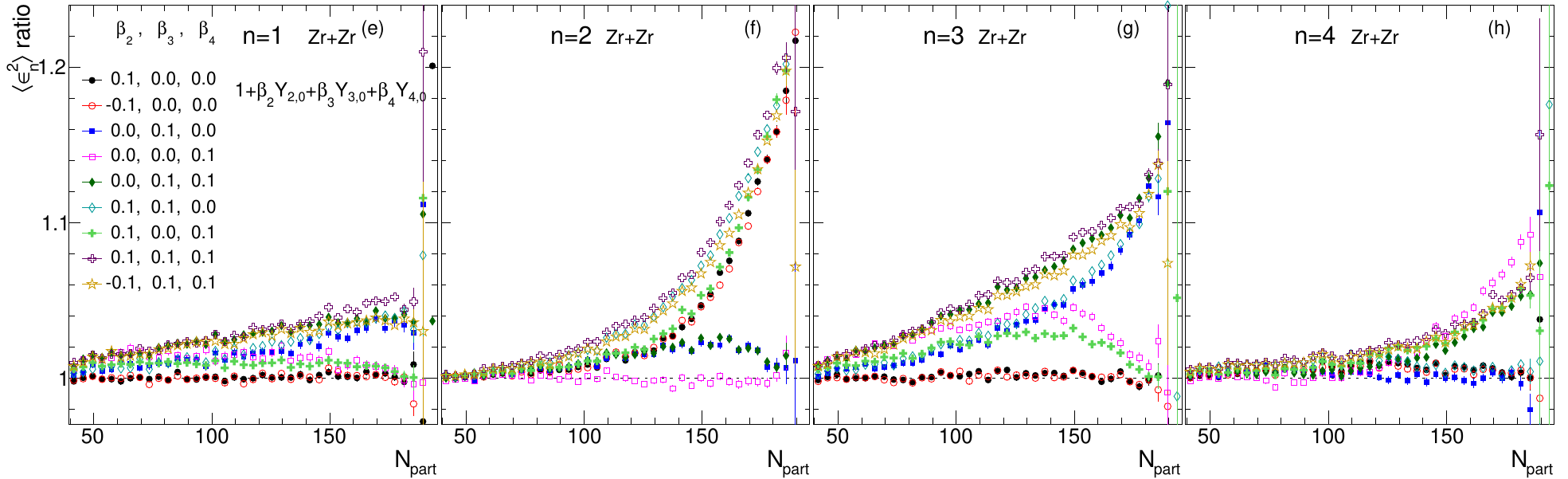}
\end{center}
\caption{\label{fig:5} Relative change of $\lr{\varepsilon_1^2}$ (left column), $\lr{\varepsilon_2^2}$ (second column), $\lr{\varepsilon_3^2}$ (third column) $\lr{\varepsilon_4^2}$ (right most column) for U+U (top row) and Zr+Zr (bottom row) collisions, relative undeformed case, for different combinations of $\beta_2$, $\beta_3$ and $\beta_4$ as indicated in the left panels.  Only the axial component of the deformation, $Y_{2,0}$,$Y_{3,0}$, and $Y_{4,0}$ are considered.}
\end{figure}

For this exploratory study, only combinations of axial-symmetric components $Y_{n,0}, n=2,3, 4$ are considered, from which the $\varepsilon_1$, $\varepsilon_2$, $\varepsilon_3$ and $\varepsilon_4$ are calculated. The analysis is carried out for different combination of $(\beta_2,\beta_3,\beta_4)$ from the values $\beta_2=\pm0.1,0$, $\beta_3=0.1,0$ and $\beta_4=0.1,0$. The results for U+U and Zr+Zr, in terms of ratios to the spherical nuclei, are shown in Fig.~\ref{fig:5}. The contributions to eccentricities from different deformation components are almost independent of each other, i.e. following Eq.~\eqref{eq:4}. Modest deviations are observed in a few cases, however. In particular, the $\varepsilon_2$ is observed to increase with $\beta_3$ in non-central region, and the difference of $\varepsilon_n$ between $\beta_2=0.1$ and $-0.1$ is also larger when $\beta_3$ and/or $\beta_4$ are non-zero. Remarkably, such non-linear effects are very small in the UCC region, where $\varepsilon_n$ is only sensitive to $\beta_n$ except for $n=1$. For the odd harmonics $\varepsilon_1$ and $\varepsilon_3$, both $\beta_3$ and $\beta_4$ can have large contribution in non-central collisions.

\begin{figure}[h!]
\begin{center}
\includegraphics[width=1\linewidth]{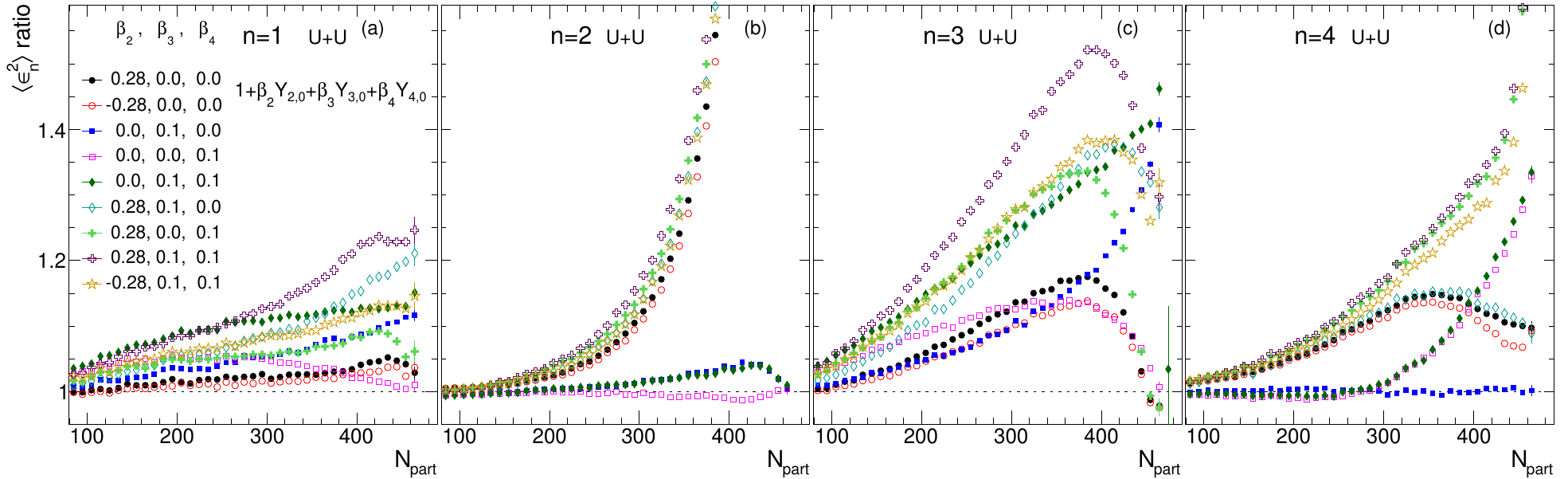}
\includegraphics[width=1\linewidth]{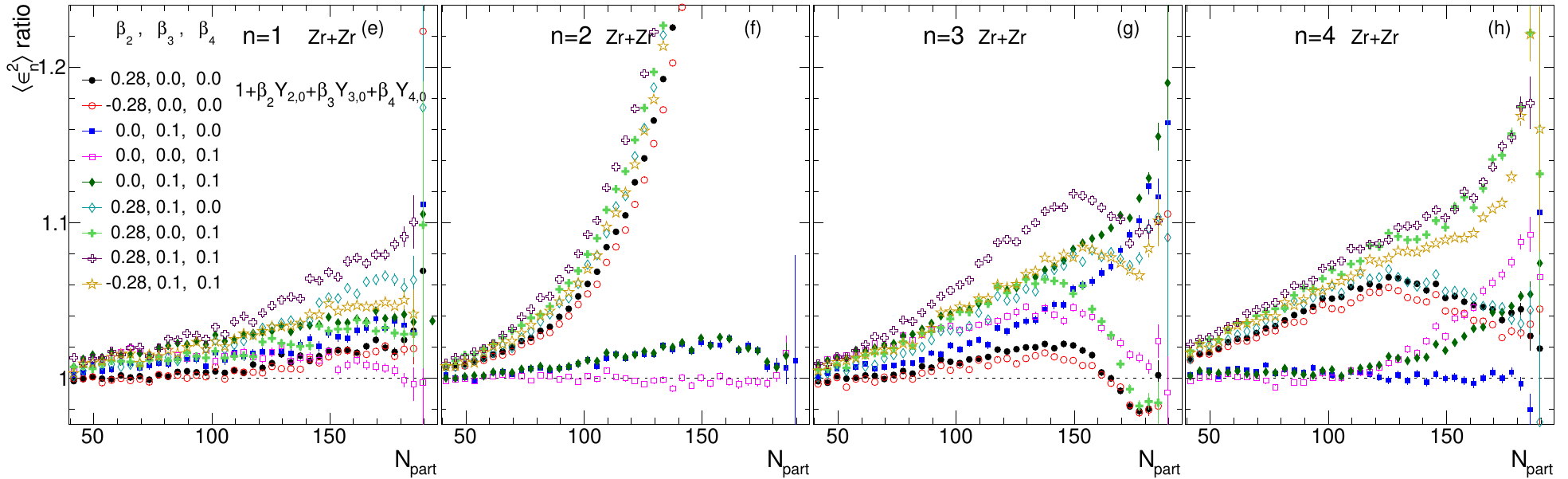}
\end{center}
\caption{\label{fig:6}  Relative change of $\lr{\varepsilon_1^2}$ (left column), $\lr{\varepsilon_2^2}$ (second column), $\lr{\varepsilon_3^2}$ (third column) $\lr{\varepsilon_4^2}$ (right most column) for U+U (top row) and Zr+Zr (bottom row) collisions, relative to undeformed case, for different combinations of $\beta_2$, $\beta_3$ and $\beta_4$ as indicated in the left panels.  Only the axial component of the deformation, $Y_{2,0}$, $Y_{3,0}$, and $Y_{4,0}$ are considered. The only difference from Fig.~\ref{fig:5} is that a larger $\beta_2$ value of 0.28 is considered.}
\end{figure}

Figure~\ref{fig:6} considers a different scenario where the quadrupole component is much larger than the octupole and hexadecapole. For this case, I increase the $\beta_2$ to the value of 0.28. Most trends remain qualitatively the same as Fig.~\ref{fig:5}. In particular, the $\varepsilon_1$ and $\varepsilon_3$ over most the centrality range, as well as $\varepsilon_4$ in the UCC region, are still dominated by the $\beta_3$ and $\beta_4$. The behaviors for $(\beta_2,\beta_3,\beta_4)=(0.28,0.1,0.1)$ for $^{238}$U, comparable to the values obtained from nuclear structure calculations in Table~\ref{tab:3}, are particularly interesting. A significant enhancement of $\lr{\varepsilon_3^2}$ of about 40\% is expected in the central collisions relative to the case of no deformation. Since $\lr{\varepsilon_3^2}\propto 1/A$ without deformation in large system, the $\lr{\varepsilon_3^2}$ in the UCC Au+Au collisions should be 238/197-1= 21\% larger than those in the UCC U+U collisions. Therefore in the presence of non-zero $\beta_3$ and $\beta_4$, the ordering is expected to be flipped: the $\lr{\varepsilon_3^2}_{\rm U}$ is expected to be 20\% larger than $\lr{\varepsilon_3^2}_{\rm Au}$, and consequently $\lr{v_3^2}_{\rm U}$ is expected to be larger than $\lr{v_3^2}_{\rm Au}$. The reverse ordering of $v_3$ between U+U and Au+Au collisions, if observed, would be a strong indication for the presence of octupole deformation in $^{238}$U nucleus. 

Another useful example is the $^{96}$Zr+ $^{96}$Zr and $^{96}$Ru+ $^{96}$Ru isobar collisions taken by the STAR Collaboration in 2018. The ratio $\lr{v_n^2}_{\rm Zr}/\lr{v_n^2}_{\rm Ru}$ will directly constrain the relative ordering of $\beta_{n,\rm Zr}$ and $\beta_{n,\rm Ru}$~\cite{Giacalone:2021uhj}, especially in the UCC region, where other effects associated with the radial distribution of nucleons, such as neutron skin, are less important~\cite{Hammelmann:2019vwd,Xu:2021vpn}. Therefore observation of significant deviation of the ratio $\lr{v_3^2}_{\rm Zr}/\lr{v_3^2}_{\rm Ru}$ from unity with the characteristic centrality dependence similar to those shown in bottom panels of Fig.~\ref{fig:5} would be a strong evidence for the presence of octupole correlations in these isobar systems.

In summary, the contributions of deformation to $\varepsilon_n$ arise mainly from $\beta_n$ for $n=2$,3 and 4. In particular, there are small cross-contributions and non-linear effects between $\beta_2$ and $\varepsilon_3$ and between $\beta_3$ and $\varepsilon_2$, especially in the UCC region. This should be contrasted to the well-known anti-correlation between $a_2'=\lr{\varepsilon_2^2}_{|\beta_n=0}$ and $a_3'=\lr{\varepsilon_3^2}_{|\beta_n=0}$ in the absence of nuclear deformation~\cite{Huo:2013qma}, where $\varepsilon_3$ arises from random position fluctuations of nucleons, the latter have opposite effects for $\varepsilon_3$ and $\varepsilon_2$. Therefore, one can constrain the value of $\beta_2$, $\beta_3$, and $\beta_4$ by combining the information from $v_1$, $v_2$, $v_3$ and $v_4$ in the ultra-central collisions as will be discussed next.

\subsection{Constraining quadrupole, octupole, and hexadecapole deformations using collective flow data}\label{sec:44}
Exploiting the linear dependence of between $\varepsilon_2^2$ and $\beta_2^2$, I have previously proposed a method~\cite{Giacalone:2021udy} to relate the quadrupole deformation between two collision systems of similar sizes. This method can be straightforwardly generalized to octupole and hexadecapole deformations.

Recall that in the UCC region, $v_n$ for $n=2-4$ depends linearly on $\varepsilon_n$, and $\varepsilon_n$ is sensitive only to $\beta_n$ according to Eq.~\eqref{eq:4}. Therefore, one expects that the $v_n$ in the UCC region to also follow a similar dependence on $\beta_n$,
\begin{equation}
\label{eq:7}
 \lr{v_n^2} =  a_n + b_n \beta_n^2,\; n=2,3,4,
 \end{equation}
where averages are performed over events in a narrow centrality class and $a_n=\lr{v_n^2}_{|\beta_n=0}$ . Following the argument of Ref.~\cite{Giacalone:2021udy}, I write down a simple equation relating the deformation and flow in two collision systems X+X and Y+Y that are close in mass number, with subscript X(Y) indicating a quantity evaluated in X+X(Y+Y) collisions:
\begin{align}
\label{eq:8}
&\beta^2_{n,\mathrm{Y}}=\left ( \frac{r_{v_n^2} r_{a,n}-1}{r_{n,\mathrm{Y}}} \right ) +\left ( r_{v_n^2}r_{b,n} \right ) \beta^2_{n,\mathrm{X}}\;,\;\;r_{v_n^2} \equiv \frac{\lr{v_n^2}_{\rm Y}}{\lr{v_n^2}_{\rm X}}\;,\; r_{b,n} = \frac{b_{n,\rm X}}{b_{n,\rm Y}}\;,\; r_{a,n} = \frac{a_{n,\rm X}}{a_{n,\rm Y}},\; r_{n,\rm Y} = \frac{b_{n,\rm Y}}{a_{n,\rm Y}}.
\end{align}

As shown in the bottom panels of Fig.~\ref{fig:1}, $b_n'$ has very weak dependence on system size, therefore one expects it is true also for $b_n$ and therefore $r_{b,n}\approx1$. In the absence of deformation $\beta_{n,\rm X}=\beta_{n,\rm Y}=0$, using the linear response relation $\lr{v_n^2} = k_n^2 \lr{\varepsilon_n^2}$, the relative difference of harmonic flow between X+X and Y+Y collisions, $\Delta \lr{v_n^2}/ \lr{v_n^2} = \left ( \lr{v_n^2}_{\rm X} - \lr{v_n^2}_{\rm Y} \right ) / \lr{v_2^2}_{\rm Y}$, can be decomposed as 
\begin{align}
\label{eq:9}
\frac{\Delta \lr{v_n^2}_{|\beta_n=0}}{\lr{v_n^2}_{|\beta_n=0}}=\frac{\Delta k_n^2}{k_n^2} +  \frac{\Delta \lr{\varepsilon_n^2}_{|\beta_n=0}}{\lr{\varepsilon_n^2}_{|\beta_n=0}} \rightarrow   \frac{\Delta a_n}{a_n} = \frac{\Delta k_n^2}{k_n^2} +  \frac{\Delta a'_n}{a'_n}\;.
\end{align}
In the UCC region, the eccentricities are dominated by the random fluctuations of nucleon positions and to a good extent can be approximated by $\lr{\varepsilon_n^2}\propto 1/A$~\cite{Alver:2010dn,Bhalerao:2011bp}, and therefore $\Delta \lr{\varepsilon_n^2}/\lr{\varepsilon_n^2} \approx \Delta \frac{1}{A}/\frac{1}{A}$. The response coefficient $k_n$ is damped with the respect to the ideal hydrodynamic value, $k_{n,{\rm ih}}$. In the simplified acoustic scaling scenario of Ref.~\cite{Gubser:2010ui,Staig:2010pn,Teaney:2012ke,Lacey:2013eia}, one has $k_n/k_{n,\mathrm{ih}}\approx 1- K n^2$, where $K$ encodes the viscous correction. This leads to $\Delta k_n/k_n \approx - \Delta K n^2 k_{n,\mathrm{ih}}/k_n$. For central collisions of large systems \cite{Alver:2010dn}, $k_n$ is close to the ideal hydro limit, and $k_{n,\mathrm{ih}}/k_n$ are nearly independent of $n$, therefore one obtains
\begin{equation} 
\label{eq:10}
\frac{\Delta k_n^2}{k_n^2} = \frac{n^2}{m^2} \frac{\Delta k_m^2}{k_m^2}\;.
\end{equation} 
Combining Eqs.~\eqref{eq:9} and \eqref{eq:10} yield two coupled equations,
\begin{equation} 
\label{eq:11}
r_{a,2}'-r_{a,2}  = x_3 (r_{a,3}'-r_{a,3}) = x_4 (r_{a,4}'-r_{a,4})\;,\;\; x_3\approx\frac{4}{9}, x_4\approx\frac{4}{16}\;.
\end{equation} 

\begin{figure}[h!]
\begin{center}
\includegraphics[width=0.32\linewidth]{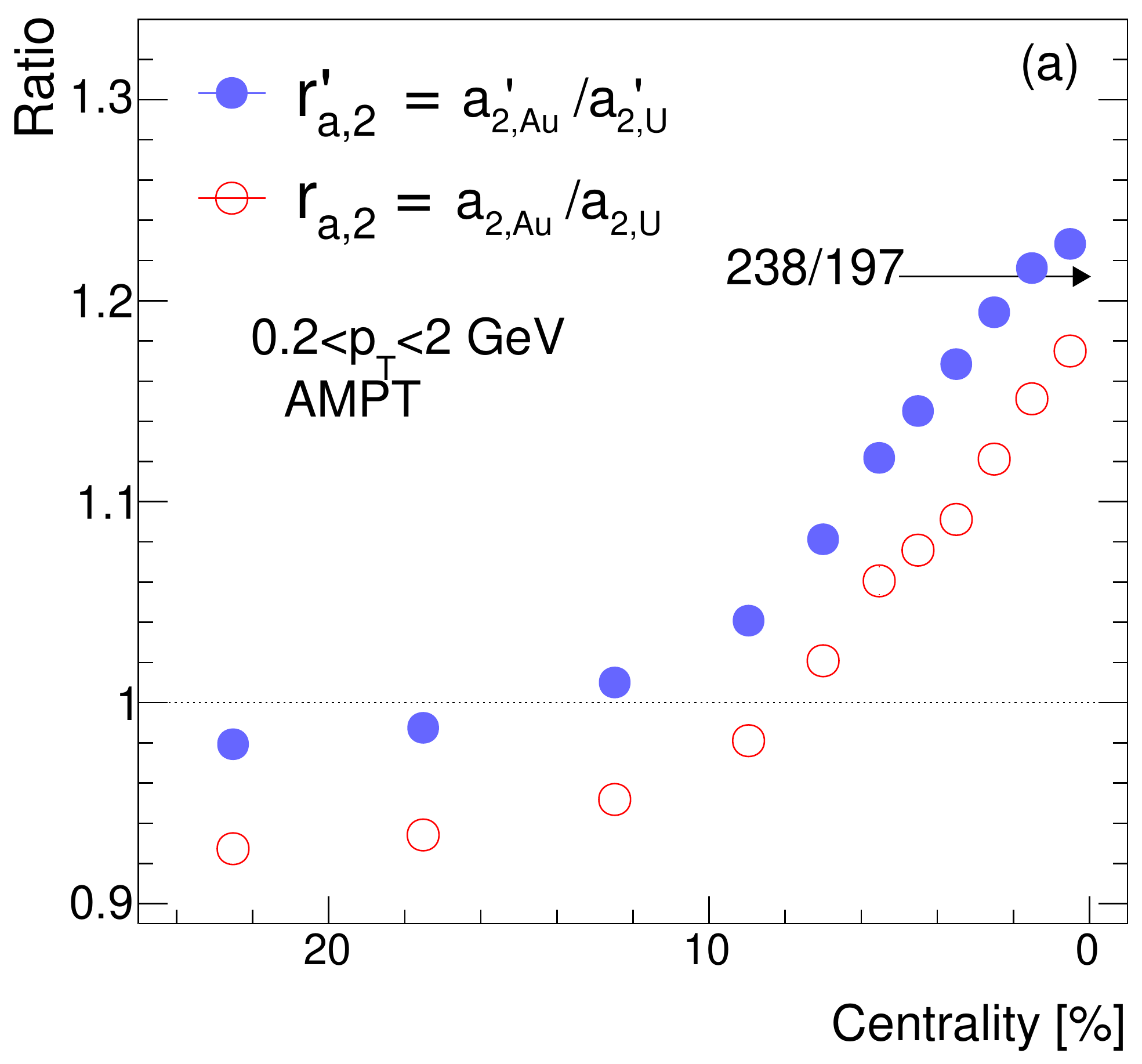}\includegraphics[width=0.32\linewidth]{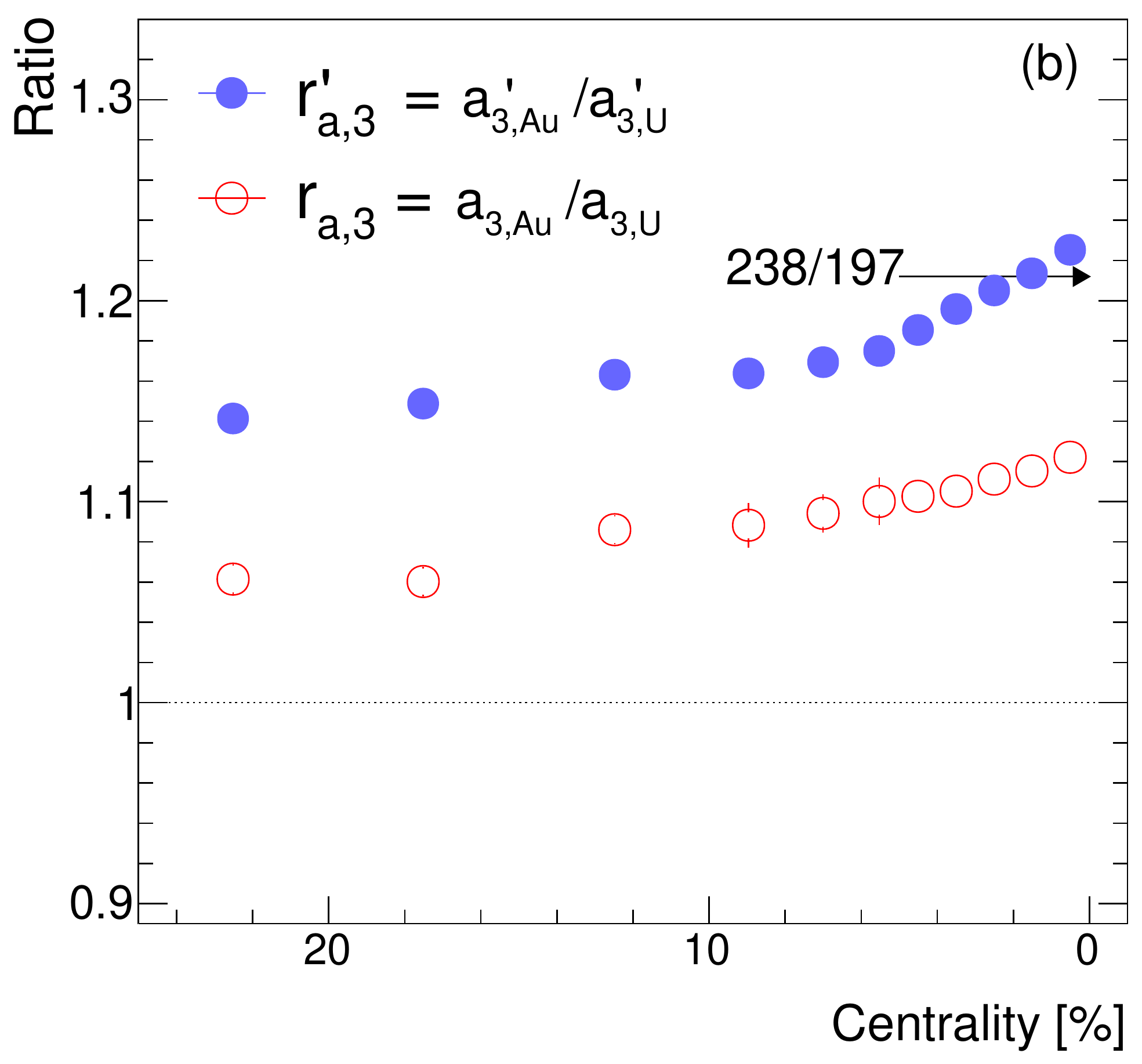}\\
\vspace*{-0.7cm}
\includegraphics[width=0.32\linewidth]{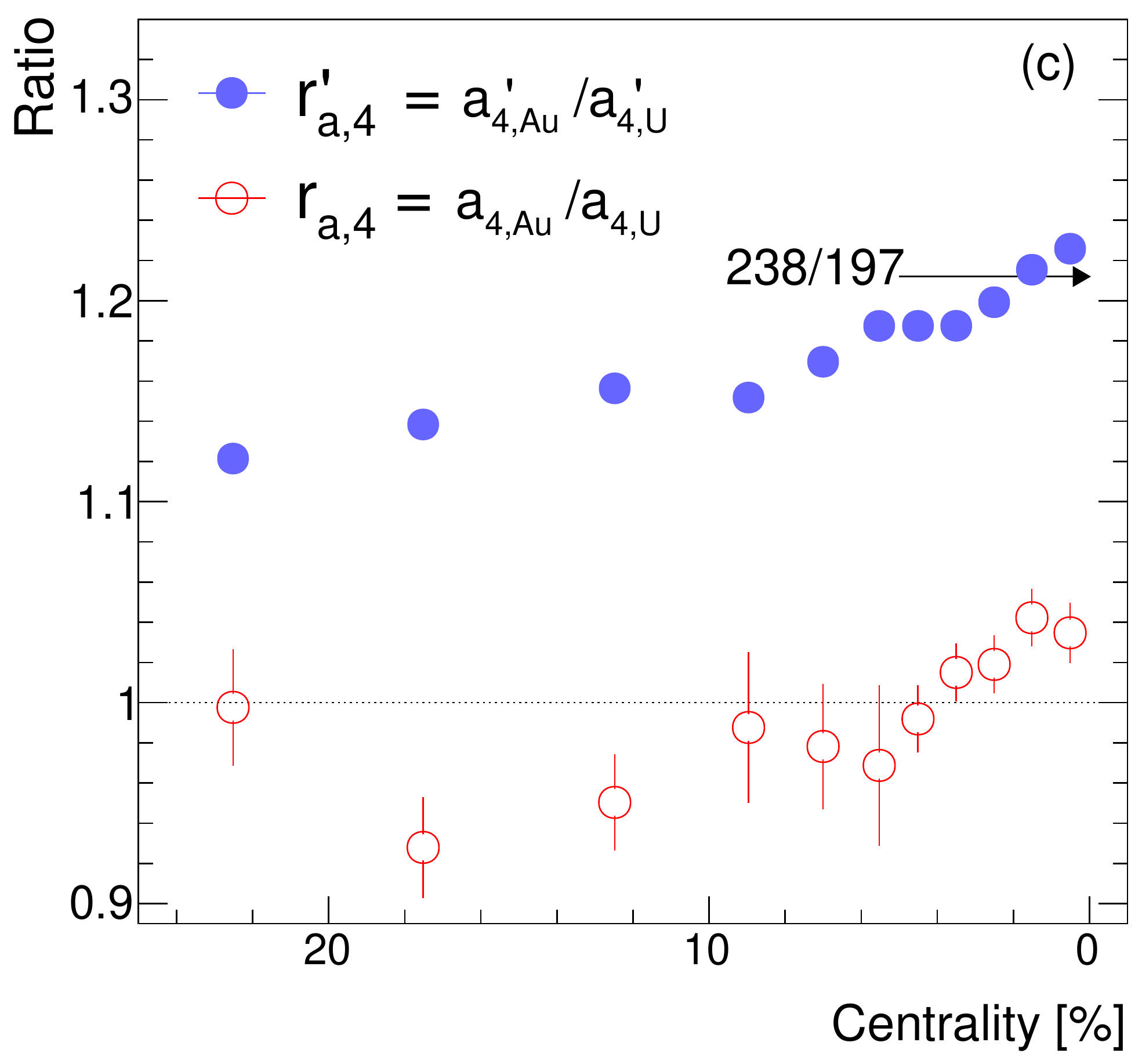}\includegraphics[width=0.32\linewidth]{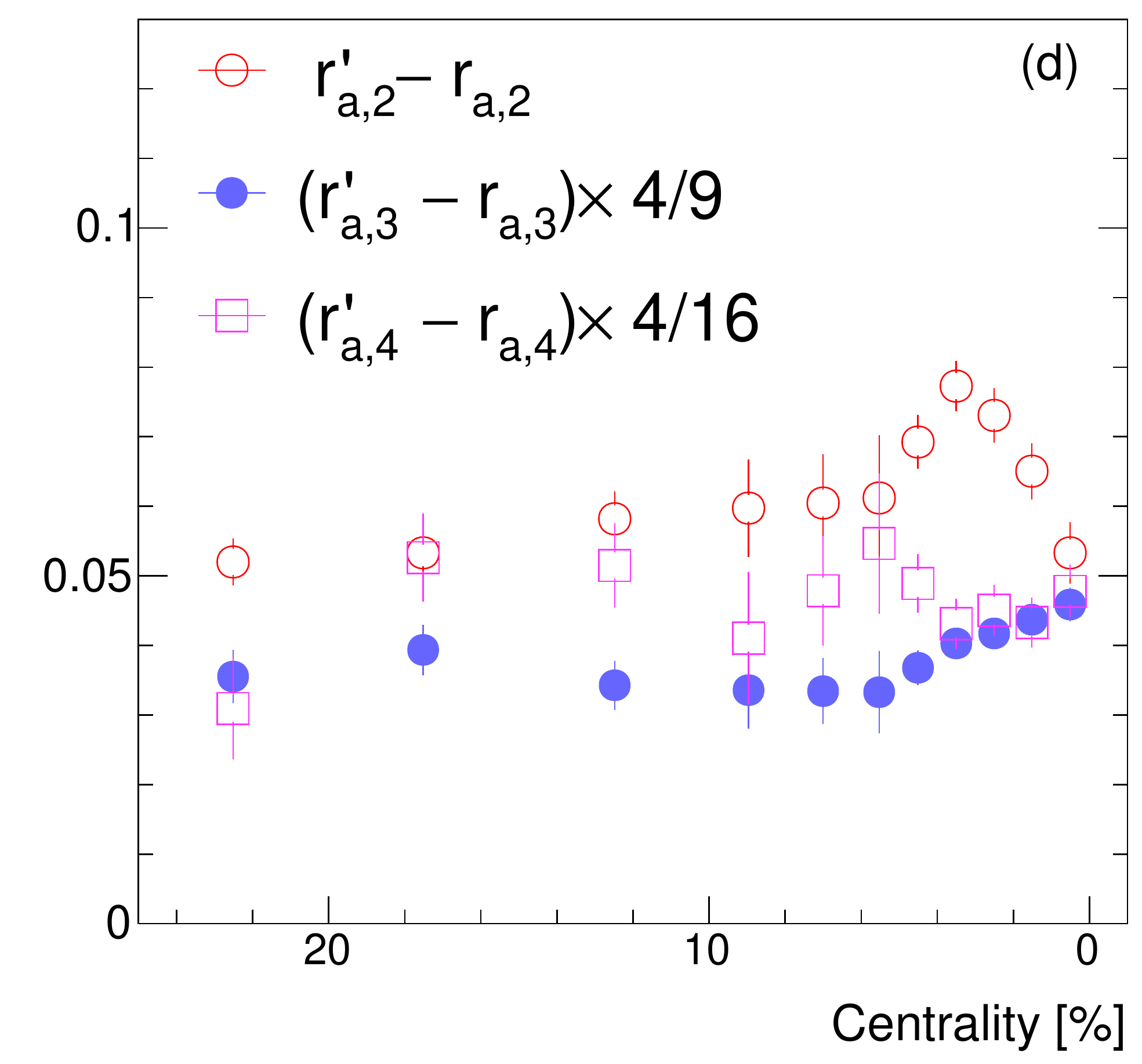}
\end{center}
\caption{\label{fig:7} The ratio of mean square eccentricity and mean square flow harmonics without deformation $a_n'=\lr{\varepsilon_n^2}_{\beta_n=0}$ and $a_n=\lr{v_n^2}_{\beta_n=0}$ between Au+Au and U+U collisions for $n=2$ (top-left), $n=3$ (top-right), $n=4$ (bottom left) as a function of centrality based on $\nch$ obtained from AMPT model in Ref.~\cite{Giacalone:2021udy}. The arrows indicate the expected ratios of ms eccentricity in the large system limit: $(1/A_{Au})/(1/A_{U})=238/197=1.21$. The bottom right panel shows the difference between $r_{a,n}'=a_{n,Au}'/a_{n,U}'$ and $r_{a,n}=a_{n,Au}/a_{n,U}$, scaled by the expected viscous damping factor according to Eq.~\eqref{eq:11}.}
\end{figure}

These equations involve only ratios of quantities between two systems close in size. All these ratios are close to unity and can be reliably estimated from the hydrodynamic model. I have verified these relations explicitly in the AMPT model simulation of Au+Au and U+U collisions without deformations in a previous study~\cite{Giacalone:2021udy,Jia:2021wbq}. The centrality dependence of these ratios are shown in Fig.~\ref{fig:7} with ${\rm X}=^{197}${\rm Au}, ${\rm Y}=^{238}${\rm U}. It is immediately clear that $r_{a,n}$ follows closely the centrality dependence trends of $r'_{a,n}$, but has smaller values due to viscous damping. The difference grows with $n$, reflecting the stronger viscous damping for higher-order flow harmonics. In the 0--1\% most central collisions (the rightmost point for each dataset), one has $r'_{a,n}=1.23$ independent of the harmonic number. This number is very close to the expected ratio of atomic numbers $(1/A_{\rm {Au}})/(1/A_{\rm{U}})=238/197=1.21$. 

The bottom right panel demonstrates the robustness of Eq.~\eqref{eq:11}. Using the viscous damping relation Eq.\eqref{eq:10}, the predicted values in the 0--1\% most central collisions converge remarkably within 0.005 between different harmonics. But deviations from this scaling appear away from the most central region. For $v_3$, one finds a value of $x_3=5/9\sim0.55$ achieves best agreements over the 1\%--25\% centrality range as shown. This value is close to $x_3=0.57 \sim 5/9$ from a recent state-of-the-art hydrodynamic simulation~\cite{Gardim:2020mmy}. For the $v_4$, I found $x_4=5/16\sim0.31$ has the best agreement in the 1\%--25\% centrality range, although the interpretation may be complicated by the mode-mixing contribution from elliptic flow that scales like $v_{4}\sim v_2^2$, for which a smaller damping $x_4=8/16$ is expected~\cite{Teaney:2012ke}. For the remaining discussion, I shall focus simply on the 0--1\% most central bin. 

First, I use the approximation $\lr{\varepsilon_n^2}_{|\beta_n=0}\propto 1/A$ and rewrite Eq.~\eqref{eq:11} and Eq.~\eqref{eq:8} as:
\begin{align}\nonumber
&\beta^2_{n,\mathrm{Y}}=\frac{r_{v_n^2} r_{a,n}-1}{r_{n,\mathrm{Y}}} + r_{v_n^2} \beta^2_{n,\mathrm{X}}\\\label{eq:12}
&r_{a,2} = (1-x_3)\frac{A_{Y}}{A_{X}} +x_3 r_{a,3} = (1-x_4)\frac{A_{Y}}{A_{X}} +x_4 r_{a,4}, \;\;x_3\approx\frac{4}{9}, x_4\approx\frac{4}{16}\;.
\end{align} 
Presumably, if one deformation e.g $n=3$ is absent, the $r_{a,3}=\lr{v_{3,X}^2}_{|\beta_3=0}/\lr{v_{3,Y}^2}_{|\beta_3=0}$ can be obtained directly from experiments, which allow us to fix $r_{a,2}$ and $r_{a,4}$ values. Alternatively, $r_{a,n}$ can be cross-calibrated by picking nuclei with similar mass number, therefore all of them are very close to unity. One such example is the Zr+Zr and Ru+Ru isobar datasets for which both $r_{b,n}$ and $r_{a,n}$ should be very close to unity~\footnote{A few percent difference in $\varepsilon_2$ might arise because the difference in neutron skin effects between $^{96}$Zr and $^{96}$Ru~\cite{Hammelmann:2019vwd,Xu:2021vpn}, but these effects are much smaller than the influence of nuclear deformation in the UCC region.}.  The only variable that needs to be evaluated numerically in hydrodynamic model is $r_{n,Y} = b_{n,Y}/a_{n,Y}$, which is the property of a single collision system. 

One such numerical analysis has been performed in Ref.~\cite{Giacalone:2021udy}, here I offer a bit more discussion on the expected behavior. Defining two response coefficients, $k_{b,n} = \sqrt{b_n/b_n'}$ and $k_{a,n} = \sqrt{a_n/a_n'}$, $r_n$ can be rewritten as 
\begin{align}\label{eq:13}
r_n =\frac{b_n}{a_n} = \frac{k_{b,n}^2}{k_{a,n}^2}\frac{b'_n}{a'_n}
\end{align}
The $k_{a,n}^2$ is the usual viscous damping coefficient for $\lr{v_n^2}$ in the absence of deformation, while $k_{b,n}^2$ describes the damping of the $\beta_n^2$-dependent part of the $\lr{v_n^2}$ in Eq.~\eqref{eq:7}. From the AMPT model study, I have found that the ratio of the two damping coefficients for $n=2$, $k_{b,2}/k_{a,2}\approx 0.75$ in the UCC region in U+U collisions, and only has a very weak dependence on $\npart$, suggesting that $k_{b,2}/k_{a,2}$ is not very sensitive to viscosity. If this is the case, the model dependence lies in the ratio $r_n'=b'_n/a'_n$, whose uncertainty arises mainly from centrality smearing effects, e.g. the relative smearing of centrality based on the final state charged particle multiplicity $\nch$ and the $\npart$. 

\section{Discussion and summary}
The main finding of the paper is the simple parametric relation between $\varepsilon_n$ and multi-pole deformation of nuclei $\beta_n$, $\lr{\varepsilon_n^2}=a'_n+b'_{n}\beta_n^2$ for $n=2$,3 and 4, valid in all centrality and different collision systems. The $a_n'$ reflects the eccentricities for spherical nuclei, i.e. $a_n'=\lr{\varepsilon_2^2}_{|\beta_n=0}$. The $a_2'$ is dominated by elliptic shape of the overlap region, which starts at a small value in central collisions and grows rapidly toward mid-central and peripheral collisions. Other $a_n', n\neq 2$ are generated by random fluctuations of participating nucleons and typically scales as $1/\npart$. On the other hand, $\beta_n$ influences the global shape of the overlap region on an event-by-event bases, and its contribution to eccentricity $b'_{n}\beta_n^2$ plays a similar role as the so called reaction plane ellipticity $\varepsilon_{2,\mathrm{RP}}$ associated with the average elliptic shape of the overlap. Due to linear response $v_n = k_n \varepsilon_n$ predicted by hydrodynamic models, I expect a similar dependence for $v_n$, $\lr{v_n^2}=a_n+b_{n}\beta_n^2$. From these, I define a deformation-dependent and deformation-independent hydrodynamic response coefficients  $k_{b,n} = \sqrt{b_n/b_n'}$ and $k_{a,n} = \sqrt{a_n/a_n'}$. It would be insightful to investigate and compare $k_{a,n}$ and $k_{b,n}$, which will provide new kind of test on the hydrodynamic models. 

\begin{figure}[h!]
\begin{center}
\includegraphics[width=0.7\linewidth]{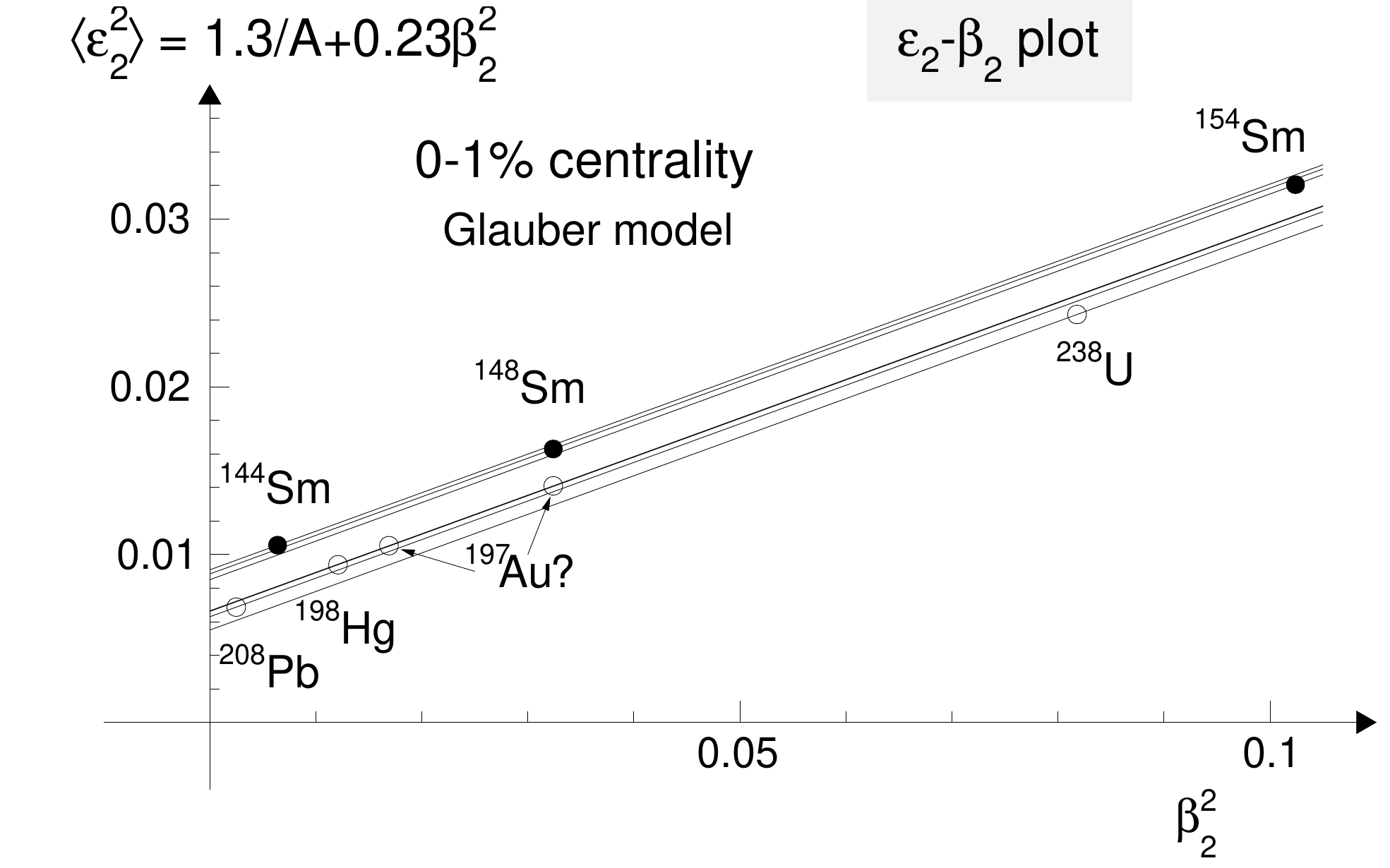}
\end{center}
\caption{\label{fig:idea2} Correlation of mean square elliptic eccentricity and $\beta_2$ in 0--1\% centrality, $\lr{\varepsilon_2^2} = a'_2+b'_2\beta_n$, estimated from Glauber model for various large collision species, where is $a'_2$ and $b'_2$ are approximately $a'_2=1.3/A$ and $b_2'=0.23$. For even-even nuclei, the values of deformation come from low-energy data, as in Table~\ref{tab:3}. For $^{197}$Au, whose $\beta_2$ is not directly measured, I also include recent estimate based on the $v_2$ data from Au+Au and U+U collisions, $|\beta_{2,\rm Au}|\approx 0.18$~\cite{Giacalone:2021udy}. Offsets between different species are due to differences in mass numbers $A$, therefore each group of nuclei with very similar masses, ($^{197}$Au, $^{198}$Hg,$^{208}$Pb) or ($^{144}$Sm, $^{148}$Sm,$^{154}$Sm), almost fall on the same curve.}
\end{figure}

The best place to reveal nuclear deformation is the ultra-central collisions (UCC) of large systems, where the deformation-driven components become comparable or even larger than the values without deformation. For this purpose, I propose a collision-system scan of a few species of similar size at RHIC to systematically establish the influence of deformation, see the sketch in Fig.~\ref{fig:idea2}. 

First, it would be useful to scan two nuclei, e.g. $^{208}$Pb and another species, in the vicinity of $^{197}$Au, to improve the modeling of Au+Au collisions, an information which is crucial for the precision interpretation of high-statistics flow data. Comparison between Pb+Pb at RHIC and the LHC will constrain any possible energy dependence of the initial state effects and pre-equilibrium dynamics. Since $^{208}$Pb is nearly spherical, a comparison of Pb+Pb with Au+Au collisions at the same energy will also allow us to better understand the impact of the moderate deformation of $^{197}$Au in Au+Au collisions. The collisions of another species e.g $^{198}$Hg+$^{198}$Hg ($\beta_2=-0.11$) would then probe more deeply the nature of the deformation of $^{197}$Au, which, being an odd-mass nucleus, hasn't been directly measured in low-energy experiments. Having additional systems also provides an independent cross-check on the initial state, for example one can setup three relations like Eq.\eqref{eq:12} to ``triangulate'' the consistency of the three deformation values. 

In the second step, one can use flow measurements in conjunction with hydrodynamics to map out the evolution of the quadrupole deformation along the chain of stable samarium isotopes. As proposed in Ref.~\cite{Giacalone:2021udy}, it would be useful to collide three isotopes: $^{144}$Sm ($\beta_2=0.08$, as spherical as $^{208}$Pb), $^{148}$Sm ($\beta_2=0.18$, triaxial much as $^{129}$Xe and $^{197}$Au), and $^{154}$Sm ($\beta_2=0.32$ well-deformed like $^{238}$U). The evolution of the quadrupole deformation can be mapped precisely at RHIC, thus offering a valuable test of nuclear structure knowledge. This scan also enables a search for enhanced octupole correlations, i.e., $\beta_{3}$ values, which are predicted to be present in the region $Z\sim 56$/$N\sim88$~\cite{Butler:2016rmu} including the samarium isotopes. The influence of octupole correlations would manifest in high-energy collisions as enhanced $v_3$, as well as modified $\rho(v_3^2,[\pT])$ correlator. Evidence of static octupole moments at low energies is rather sparse, and heavy ion collisions might be a more sensitive approach. 

In summary, I have studied the parametric dependence of eccentricity $\varepsilon_n$ on the quadrupole $\beta_2$, octupole $\beta_3$ and hexadecapole $\beta_4$ deformation of the nucleus in heavy ion collisions. The mean square eccentricity $\lr{\varepsilon_n^2}$ are found to depend linearly on $\beta_n^2$ for $n$=2, 3 and 4. I also find that $\beta_3$ contributes significantly to $\varepsilon_1$, and in non-central collisions, the $\beta_4$ also contributes to $\varepsilon_1$ and $\varepsilon_3$. In central collisions, there are very little cross contribution between $\beta_2$ and $\varepsilon_3$ and between $\beta_3$ and $\varepsilon_2$, although in non-central collisions, $\beta_3$ ($\beta_2$) contributes modestly to $\varepsilon_2$ ($\varepsilon_3$). Since harmonic flow $v_n$ is directly driven by the corresponding eccentricity, $v_n\propto\varepsilon_n$, one expects very similar parametric dependencies between $v_n$ and $\beta_m$ for both $m=n$ and $m\neq n$. These findings provide a strong motivation to use heavy ion collisions as a precision tool to scan and map out the ground state nuclear deformations and compare with low energy nuclear measurements, not only for the quadrupole deformation, but also for the octupole deformation, whose evidence is quite sparse in nuclear structure physics. The procedure for doing this is presented. One finds that the non-axial deformations, in particular the triaxiality of quadrupole deformation, do not influence $\lr{\varepsilon_n^2}$, but they can be probed by other observables such as $v_n-[\pT]$ correlation, and possibly higher-order cumulants of $v_n$ and $[\pT]$. The information about the shape of atomic nuclei obtained in heavy-ion collisions is fully complementary to that obtained in nuclear structure experiments. A carefully planned system scan of stable species in the nuclear chart at RHIC, the LHC, and other collider facilities could open new direction of research in nuclear physics. 

{\bf Acknowledgements:} I thank Giuliano Giacalone for stimulating discussions and recent collaborations that motivated this work. I thank Shengli Huang and Chunjian Zhang for valuable discussions. This work is supported by DOE DEFG0287ER40331 and NSF PHY-1913138.

\appendix
\section{An analytical estimate in head-on collisions}\label{sec:app1}
I consider liquid-drop nuclear potential with a sharp surface, $\rho(r,\theta,\phi)=\rho_0$ when $r<R(\theta,\phi)$ and zero otherwise, and focus on head-on collisions with maximum overlap, i.e the two nuclei not only have zero impact parameter, but also need to be aligned in a way to ensure the overlap region contains all the nucleons $\npart=2A$. However, at the end of this section, I also relax the requirement to consider only zero impact parameter, which corresponds more closely to the realistic scenario when the effects of centrality resolution is taken into account. Our goal is to establish the simple parametric dependence between $\lr{\varepsilon_n^2}$ and $\beta_m$ in Eq.~\eqref{eq:4}. 

For this discussion, I switch back to using complex spherical harmonics $Y_n^m$ with the normalization condition on the coefficients $\sum_{m=-n}^{n} |\alpha_{n,m}|^2=1$ and $\alpha_{n,m} = \alpha_{n,-m}^{*}$. Let's first re-express Eq.~\eqref{eq:3} as an integral in the 3D, using the relation $r_{\perp}=r\sin\theta$:
\begin{align}\label{eq:3b}
\varepsilon_ne^{in\Phi_n} = - \frac{\int  r^n \sin^n\!\theta e^{in\phi} \rho(\vec{r}) d^3\vec{r}}{\int  r^n \sin^n\!\theta \rho(\vec{r}) d^3\vec{r}}= -\sqrt{\frac{4\pi (2n)!!}{(2n+1)!!}}\frac{\int (1+\sum_{l,m} \beta_l \alpha_{l,m}Y_l^m)^{n+3} Y_n^n  \sin\!\theta d\theta d\phi }{\int (1+\sum_{l,m} \beta_l \alpha_{l,m}Y_l^m)^{n+3} \sin^{n+1}\!\theta d\theta d\phi},
\end{align}
where I have used $Y_n^n=\sqrt{\frac{(2n+1)!!}{4\pi (2n)!!}}\sin^n\!\theta e^{in\phi}$ and the fact that the range of integration along radial direction is $r\in[0,R_0(1+\sum_{l,m}\beta_l\alpha_{l,m}Y_{l}^m)]$. Keeping the integration to leading order in $\beta_n$, one has 
\begin{align}\label{eq:3c}
\varepsilon_ne^{in\Phi_n} = -A_n\int (\sum_{l,m} \beta_l \alpha_{l,m}Y_l^m Y_n^n) \sin\!\theta d\theta d\phi = -A_n \beta_n \alpha_{n,n},\; A_n\equiv \frac{(n+3)\Gamma(1+1/2+n/2)}{\pi\Gamma(1+n/2)}\sqrt{\frac{(2n)!!}{(2n+1)!!}}
\end{align}
This result is easy to understand, for tip-tip collision where the $z$-axis is aligned with beam direction (middle row of Fig.~\ref{fig:idea}), only the $Y_{n}^n$ component can contribute to the eccentricity. The results, $\varepsilon_n = A_n\beta_n/\sqrt{2}$, for nuclear surface containing only this mode $R=R_0(1+\beta_nY_{n,n})=R_0(1+\beta_n/\sqrt{2}(Y_{n}^{-n}+(-1)^nY_{n}^{n}))$, are listed as the first term of each entry in the second row of Table~\ref{tab:2}.

In order to calculate the body-body collision for $Y_{n,0}=Y_n^0$ shown in top row of Fig.~\ref{fig:idea}, the nuclear surface, or equivalently the direction of projection $Y_n^n$, need to be rotated by Euler angles $(\alpha_e,\beta_e,\gamma_e)=(0,\pi/2,0)$, i.e. $Y_n^n=\sum_{m'}D_{n,m'}^n(0,\pi/2,0) Y_n^{m'}$, where $D_{n,m'}^n$ is the Wigner rotational matrix. Plugging this into Eq.~\eqref{eq:3c} and considering only axial deformation $R=R_0(1+\beta_nY_{n}^0)$, give, $\varepsilon_n = A_n D_{n,0}^n(0,\pi/2,0)\beta_n=\sqrt{(2n)!}/(n!2^n) A_n\beta_n$. The values are provided as the first term of each entry in the top row of Table~\ref{tab:2}.

The calculation of Eq.~\eqref{eq:3b} including higher-order terms in $\beta_n$ is straightforward; and the full expression up to the second order expansion are listed in the table. Interestingly, It is found that if one only considers the expansion for nuclear surface described by axial deformations $R=R_0(1+\beta_2Y_{2,0}+\beta_3Y_{3,0}+\beta_4Y_{4,0})$ in the numerator of an equation similar to Eq.~\eqref{eq:3b}, the ratios of the coefficients of the high-order terms to that of the leading order in the first row of Table~\ref{tab:2} are exactly the same as the Eq.~(13) in Ref.~\cite{Ryssens:2018dza}. However, including also the influence of the denominator of Eq.~\eqref{eq:3b} modifies the coefficients of some of the second-order terms.

Next, I consider random orientation of the nucleus around its center of mass described by Euler angles $\Omega=(\alpha_e,\beta_e,\gamma_e)$, and the rotations of the two nuclei are required to be the same. For the purpose to calculating the $\varepsilon_n$, it is equivalent to rotate the transverse plane by applying the substitution, $Y_n^n\rightarrow \sum_{m'} D_{n,m'}^n(\alpha,\beta,\gamma) Y_n^{m'}$, in Eq.~\eqref{eq:3c},
\begin{align}\label{eq:3d}
\varepsilon_ne^{in\Phi_n} = -A_n\int (\sum_{l,m,m'} \beta_l \alpha_{l,m} D_{n,m'}^n Y_l^m Y_n^{m'}) \sin\!\theta d\theta d\phi=-A_n (\sum_{l,m,m'} \beta_l \alpha_{l,m}  D_{n,m'}^n \delta_{ln} \delta_{m,-m'})=-A_n \beta_n \sum_m \alpha_{n,m}  D_{n,m}^n
\end{align}
Using the orthogonality relation for D-matrix, the mean square average over Euler angle is 
\begin{align}\label{eq:3e}
\lr{\varepsilon_n^2} =A_n^2\beta_n^2 \int (\sum_m  \alpha_{n,m}  D_{n,m}^n)(\sum_{m'} \alpha_{n,m'}  D_{n,m'}^n)^* \frac{d\Omega}{8\pi^2}=\frac{A_n^2}{2n+1}\beta_n^2.
\end{align}
The result apparently is independent of the mixture of different shape component $Y_n^m$, as long as the overall magnitude $\beta_n$ remains the same. This numerical values are listed in the third row of Table~\ref{tab:2}.

The dipolar eccentricity can be calculated in a similar way. The angular weight involved is decomposed into two spherical harmonics, $\sin^3\!\theta e^{i\phi} =8/5\sqrt{\pi/21}(Y_3^1-\sqrt{14}Y_1^1)$, which implies that a octupole nuclear shape can give rise to a dipole eccentricity. As the original nuclear surface has no dipole component, $Y_1^1$ drops out, and Ione obtains,
\begin{align}\label{eq:3f}
\varepsilon_1e^{i\Phi_1} = - \frac{\int  r^3 \sin^3\!\theta e^{i\phi} \rho(\vec{r}) d^3\vec{r}}{\int  r^3 \sin^3\!\theta \rho(\vec{r}) d^3\vec{r}} \approx -\frac{64}{5\sqrt{21\pi^3}}\beta_3\int \sum_m\alpha_{3,m}Y_3^m (\sum_{m'}D_{1,m'}^3Y_3^{m'})=-\frac{64}{5\sqrt{21\pi^3}}\beta_3 \sum_{m}\alpha_{3,m}D_{1,m}^3.
\end{align}
Therefore the ms average of dipolar eccentricity over all orientations gives
\begin{align}\label{eq:3g}
\lr{\varepsilon_1^2} =\frac{64^2}{525\pi^3}\beta_3^2\frac{1}{8\pi^2}\int \sum_{m,m'}\alpha_{3,m}\alpha_{3,m'}^{*}D_{1,m}^3D_{1,m}^{3*} d\Omega = \frac{4096}{3625\pi^3}\beta_3^2,
\end{align}
again independent of the mixture of different shape component $Y_3^m$. Note that, this remarkable contribution is present entirely because the $r_{\perp}^3$ weight in the definition of $\varepsilon_1$ in Eq.~\eqref{eq:3}, which is naturally required by the cumulant framework~\cite{Teaney:2010vd}.

Next, let us consider the possible contribution of quadrupole deformation to the $\varepsilon_4$. For this purpose, one expands the nuclear shape in the numerator of Eq.~\eqref{eq:3b} and keeping terms that are proportional to $\beta_2^2$,
\begin{align}\nonumber
\varepsilon_4 &\approx 3A_4\beta_2^2\int \sum_{m,m'}\alpha_{2,m}\alpha_{2,m'} D_{4,-m-m'}^4 Y_2^m Y_2^{m'}Y_4^{-m-m'} \sin\!\theta d\theta d\phi\\
&=  3A_4\beta_2^2 \frac{15}{\sqrt{4\pi}}  \sum_{m,m'}\alpha_{2,m}\alpha_{2,m'} D_{4,-m-m'}^4\left(\begin{array}{ccc}
2 & 2 & 4 \\
0 & 0 & 0
\end{array}\right)\left(\begin{array}{ccc}
2 & 2 & 4 \\
m & m' & -m-m'
\end{array}\right)\\\label{eq:3fb} 
&= \frac{45}{2\pi}\sqrt{\frac{2}{35}} \beta_2^2[\alpha_{4,0}' D_{4,0}^4+\frac{\alpha_{4,2}'}{\sqrt{2}} (D_{4,2}^4+ D_{4,-2}^4)+\frac{\alpha_{4,4}'}{\sqrt{2}} (D_{4,4}^4+ D_{4,-4}^4)]
\end{align}
where $\alpha_{4,0}'=(7+5\cos(2\gamma))/12$, $\alpha_{4,2}'=\sqrt{5/12}\sin(2\gamma)$ and $\alpha_{4,4}'=(1-\cos(2\gamma))\sqrt{35}/12$, satisfying $(\alpha_{4,0}')^2+(\alpha_{4,2}')^2+(\alpha_{4,4}')^2=1$. The average over Euler angles gives
\begin{align}\label{eq:3gb} 
\lr{\varepsilon_4^2} =\frac{45^2}{4\pi^2}\frac{2}{35}\beta_2^4 \int (\sum_m  \alpha_{4,m}'  D_{4,m}^4)(\sum_{m'} \alpha_{4,m'}'  D_{4,m'}^4)^* \frac{d\Omega}{8\pi^2}=\frac{45}{14\pi^2}\beta_2^4\;.
\end{align}

Lastly, I consider the case when one only keeps the zero impact parameter requirement. Ignoring the few nucleons (or a small portion of the volume) that may not in the overlap region, Eq.~\eqref{eq:3d} is simply the average of two nuclei with different Euler angles $\Omega_1$ and $\Omega_2$~\footnote{I am unable to derive an analytical formula for the contribution of the small portion of volume not included in the overlap region. For a few special cases, not including this volume are found to reduce slightly the coefficients of the $\beta_n^2$ dependence.}
\begin{align}\label{eq:3h}
\varepsilon_ne^{in\Phi_n} = -A_n \frac{\beta_n}{2} \sum_m \alpha_{n,m} (D_{n,m}^n(\Omega_1)+ D_{n,m}^n(\Omega_2))
\end{align}
From this, the mean square average similar to Eq.\eqref{eq:3e} needs to be integrated over both $\Omega_1$ and $\Omega_2$. The crossing terms such as $D_{n,m}^n(\Omega_1)D_{n,m}^n(\Omega_2)^*$ vanish after this integration, and the final result is exactly half of the original value. This argument also applies to the $\lr{\varepsilon_1^2}$ in Eq.\eqref{eq:3g} and the quadrupole contribution to $\varepsilon_4$ in Eq.~\eqref{eq:3gb}. So for a more realistic selection of ultra-central collisions corresponding to close to zero impact parameter, the coefficients of the $\beta_n^2$ dependence are a factor of two smaller. These values are listed in the bottom row of Table~\ref{tab:2} and they are closer to Monte Carlo Glauber model result shown in the bottom row of Figs.~\ref{fig:1} and \ref{fig:2}.

\section{More detailed results}\label{sec:app2}
\begin{figure}[h!]
\begin{center}
\includegraphics[width=1\linewidth]{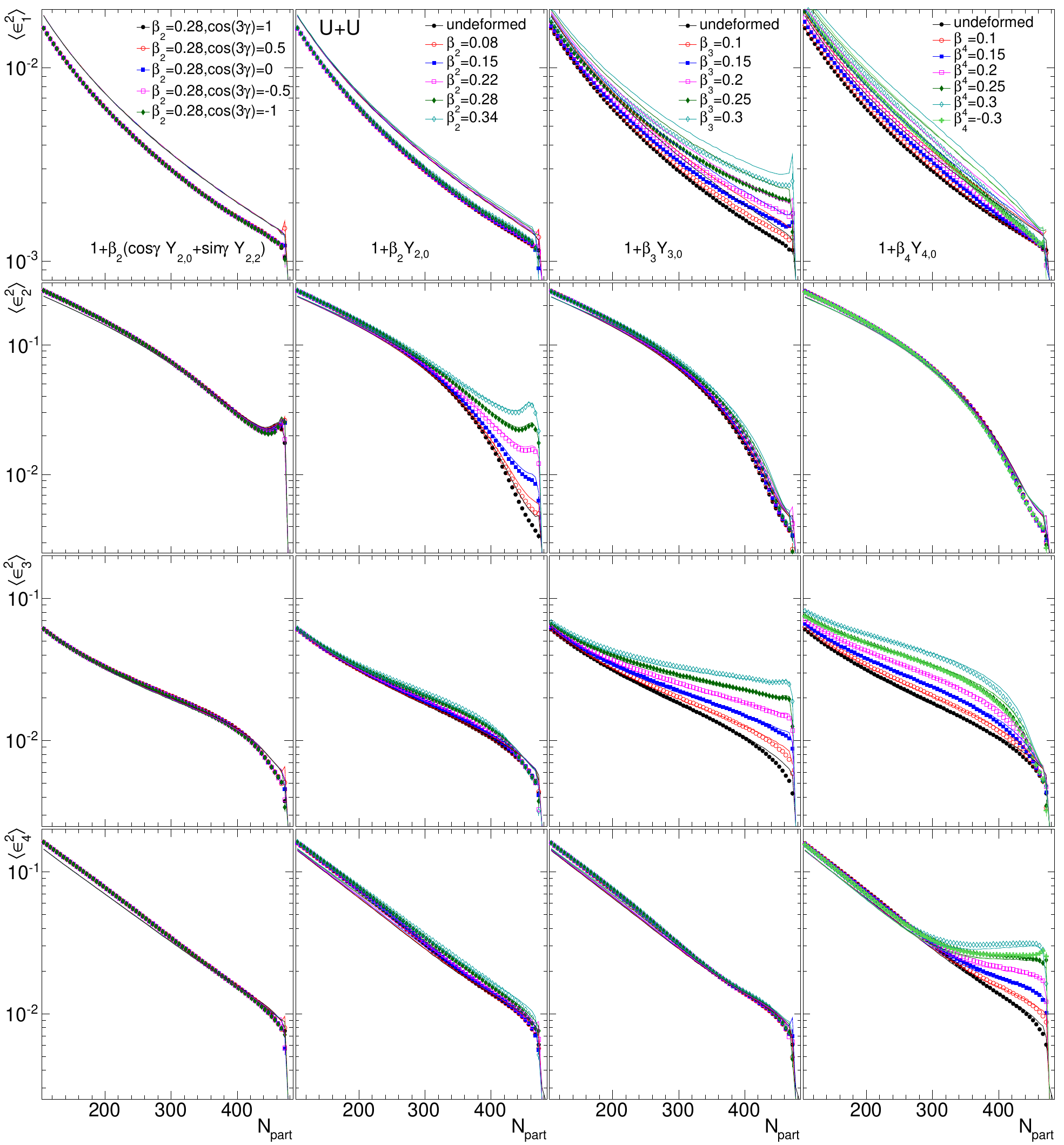}
\end{center}
\caption{\label{fig:app0}  The centrality dependence, characterized by $\npart$, of $\lr{\varepsilon_n^2}$ in U+U collisions in the presence of different quadrupole triaxiality $\gamma$ for $\beta_2=0.28$ (left column), different axial quadrupole deformation $\beta_2$ (second column), different axial octupole deformation $\beta_3$ (third column) and different axial hexadecapole deformation $\beta_4$ (last column) for $n=1$ (top row), $n=2$ (second row), $n=3$ (third row) and $n=4$ (bottom row). The markers and line curves represent $\lr{\varepsilon_n^2}$ calculated from nucleon Glauber model and quark Glauber model, respectively. The functional form of the deformation and different parameters are given in the top-row panel for each corresponding column.}
\end{figure}

\begin{figure}[h!]
\begin{center}
\includegraphics[width=1\linewidth]{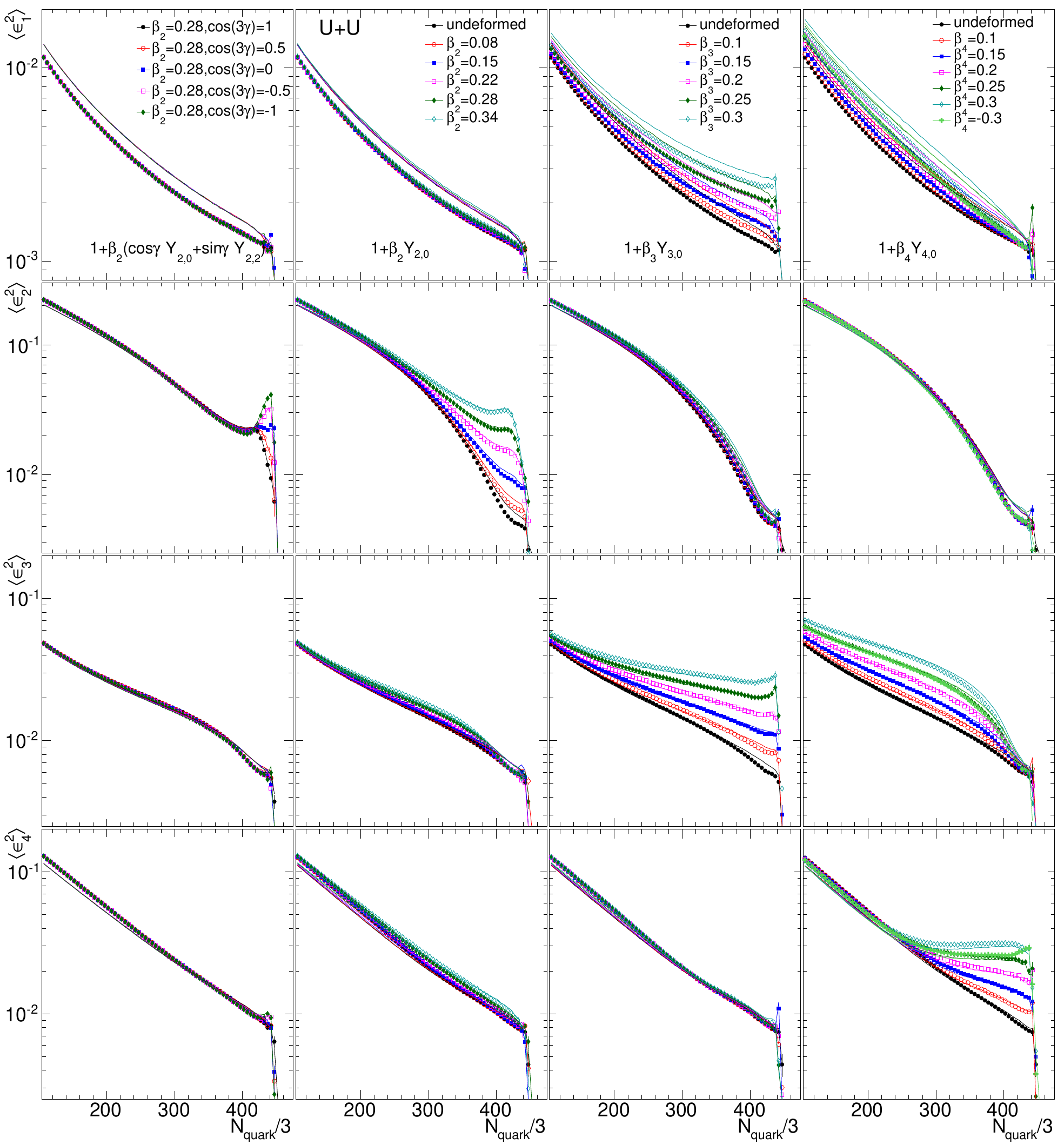}
\end{center}
\caption{\label{fig:app1}  Same as Fig.~\ref{fig:app0} but using the $\nqp$ as centrality.}
\end{figure}

\begin{figure}[h!]
\begin{center}
\includegraphics[width=1\linewidth]{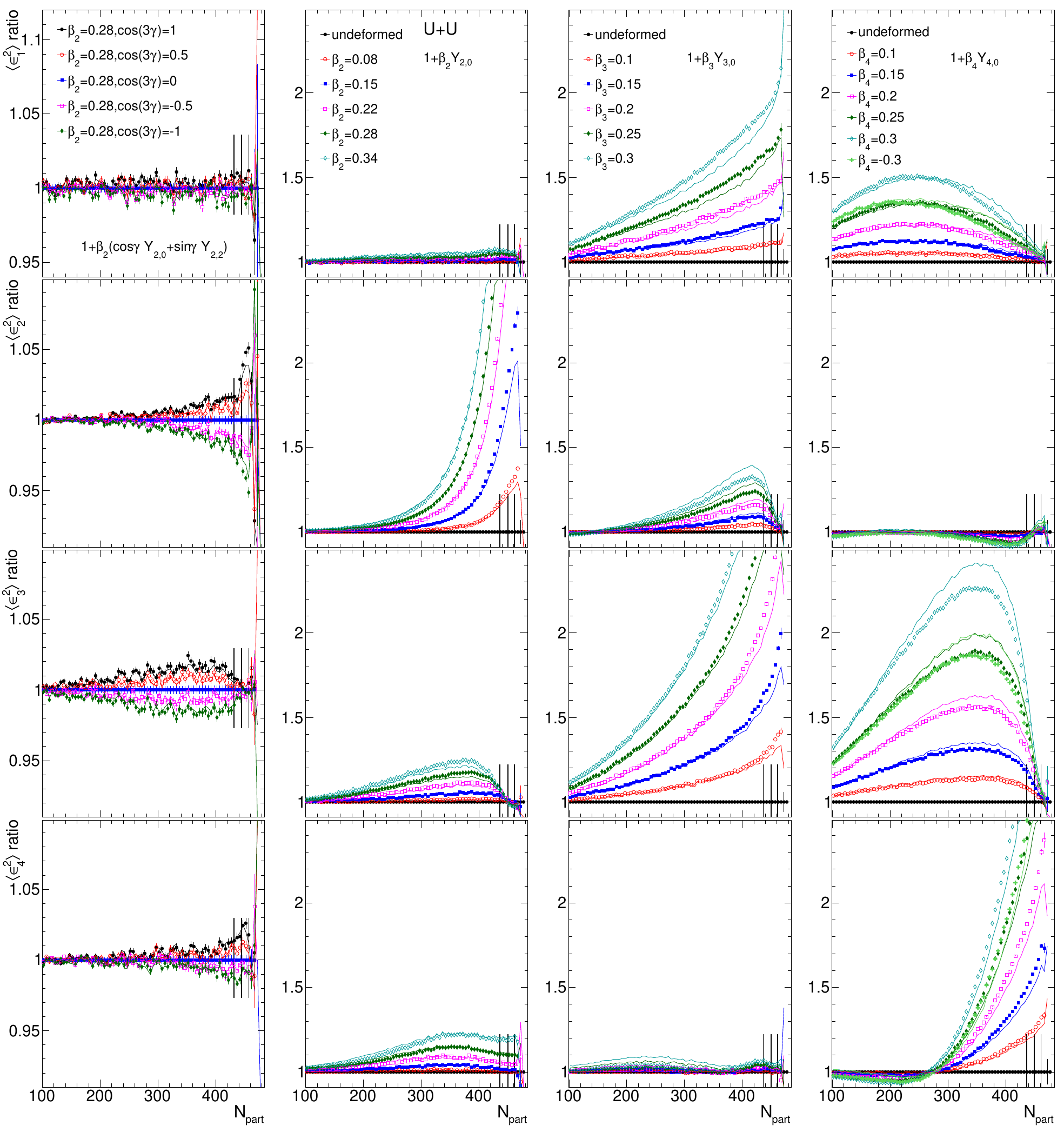}
\end{center}
\caption{\label{fig:app2} The ratio of $\lr{\varepsilon_n^2}$ to the $\lr{\varepsilon_n^2}$ in U+U collisions calculated with the default choice of each parameter (indicated by marks or lines around unity in each panel); they are obtained directly from Fig.~\ref{fig:app0}. The three solid vertical bars around unity in each one of these ratio plots indicate the locations of 2\%, 1\% and 0.2\% centrality.}
\end{figure}

\begin{figure}[h!]
\begin{center}
\includegraphics[width=1\linewidth]{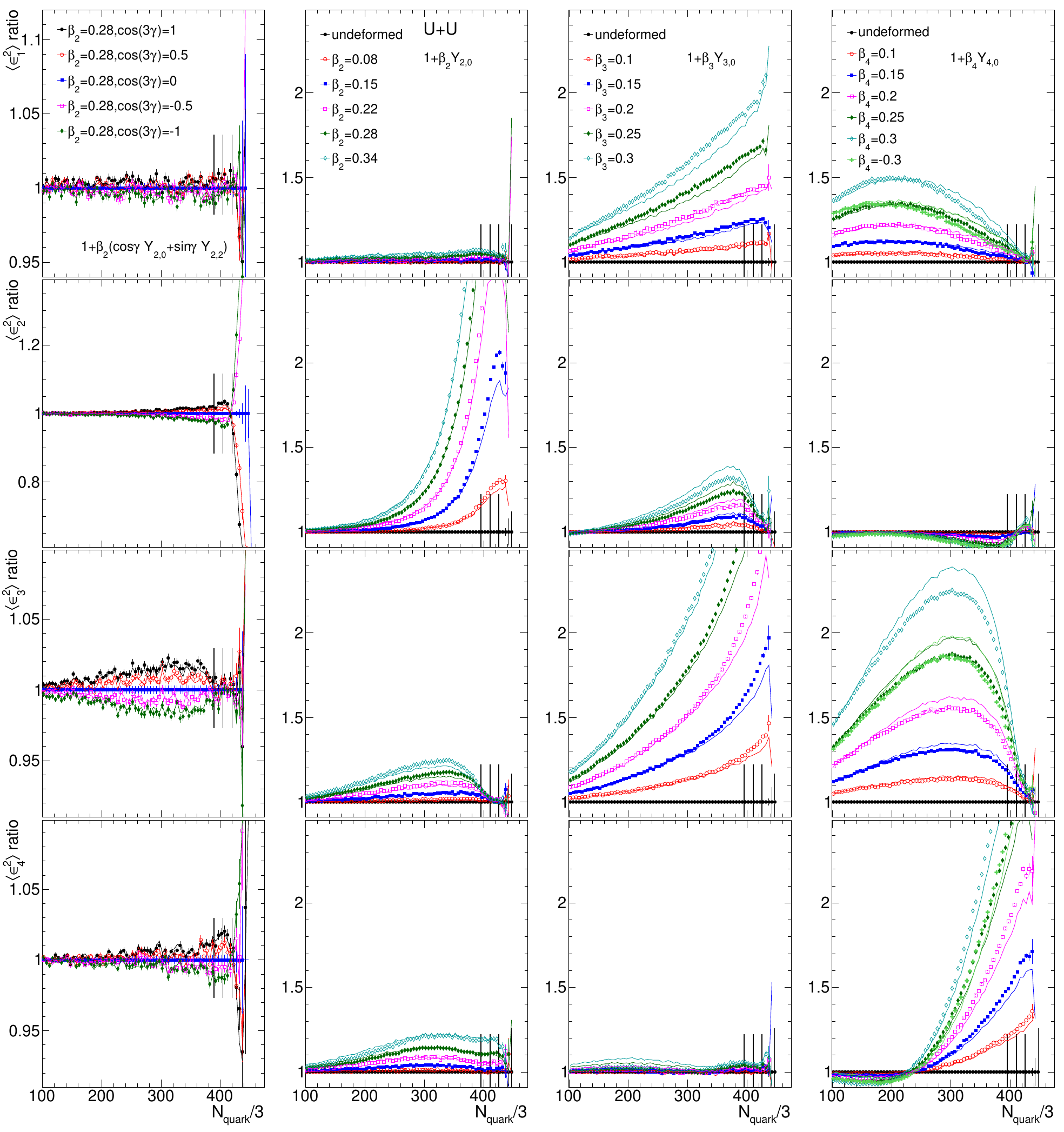}
\end{center}
\caption{\label{fig:app3}  Same as Fig.~\ref{fig:app2} but using the $\nqp$ as centrality.}
\end{figure}

\begin{figure}[h!]
\begin{center}
\includegraphics[width=1\linewidth]{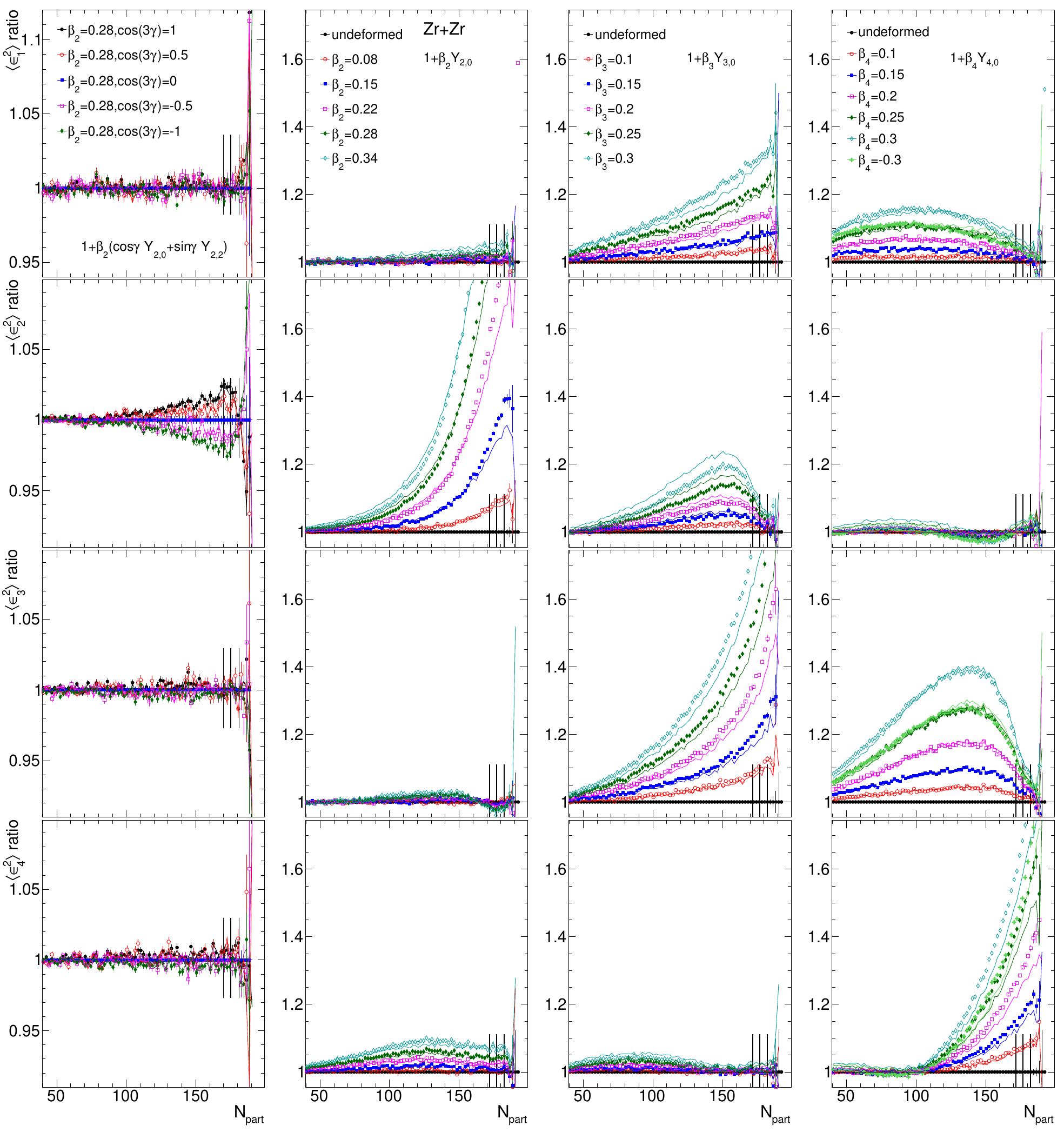}
\end{center}
\caption{\label{fig:app2b}  Same as Fig.~\ref{fig:app2} but for Zr+Zr collisions.}
\end{figure}

\begin{figure}[h!]
\begin{center}
\includegraphics[width=1\linewidth]{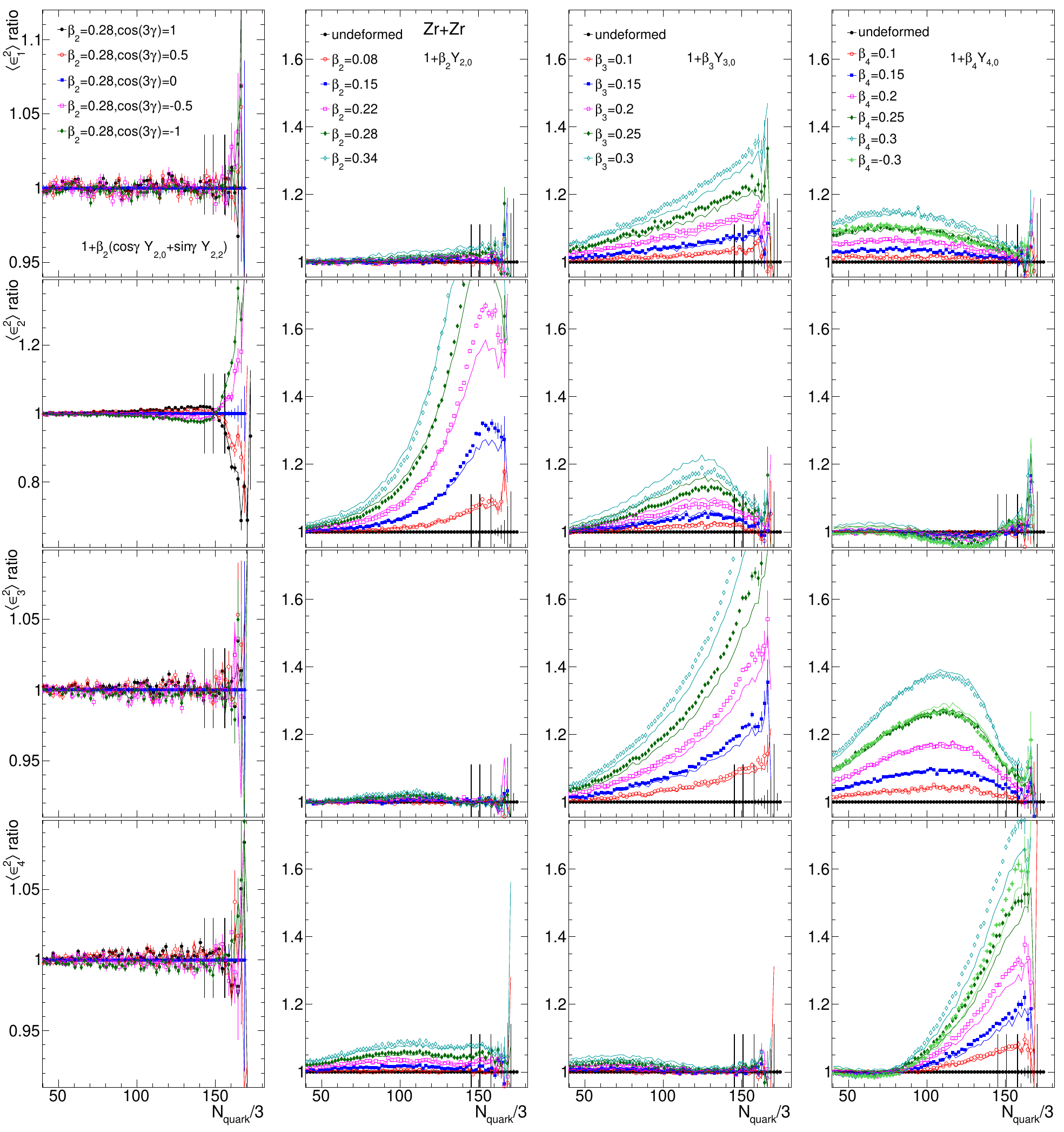}
\end{center}
\caption{\label{fig:app3b}  Same as Fig.~\ref{fig:app3} but for Zr+Zr collisions.}
\end{figure}

For completeness, the full set of correlations between mean square eccentricities $\lr{\varepsilon_n^2}$ and deformation parameters, $\gamma$, $\beta_2$, $\beta_3$ and $\beta_4$ in Eq.~\eqref{eq:1}, are included here. The $\lr{\varepsilon_n^2}$ are calculated using the nucleon Glauber model and quark Glauber model in the U+U collisions. They are compiled in Figs.~\ref{fig:app0} and \ref{fig:app1} as a function of two centrality estimators, $\npart$ and $\nqp$, respectively. From these plots, the ratios of $\lr{\varepsilon_n^2}$ to that obtained for default choice of each parameter are plotted in Figs.~\ref{fig:app2} and \ref{fig:app3}, respectively. For the smaller Zr+Zr collision system, I only show the ratios in Figs.~\ref{fig:app2b} and \ref{fig:app3b}.

The main difference between the two centrality estimators is in the behavior $\lr{\varepsilon_n^2}$ in the UCC region, more clearly visible in the ratio plots. However, whether $\lr{\varepsilon_n^2}$ themselves are calculated from nucleons or quarks have little influences on these ratios. Another important point is about the contribution of $\beta_m$ to $\lr{\varepsilon_n^2}$ for $m\neq n$. Although such mixings could in principle be used to constrain the $\beta_3$ using $v_1$ as well as $\beta_4$ using $v_1$ and $v_3$, this mixing also forbiddens a straightforward disentanglement of different deformation components. Fortunately, such mixing effects are minimal in the UCC collisions, and if one stay in the 0--1\% centrality range, each $\lr{\varepsilon_n^2}$ only has one dominating contribution: $\lr{\varepsilon_n^2}_{\rm{UCC}}=a_n'+b_{n}'\beta_n^2$ for $n=2$, 3 and 4, and $\lr{\varepsilon_1^2}_{\rm{UCC}}=a_1'+b_{3,1}'\beta_3^2$. In non-central collisions, there is a modest cross correlation between $\beta_3$ and $\lr{\varepsilon_2^2}$ and between $\beta_2$ and $\lr{\varepsilon_3^2}$ with the former having somewhat larger amplitudes. However, in medium-size Zr+Zr collision system, the cross correlation between $\beta_3$ and $\lr{\varepsilon_2^2}$ is still significant, but almost disappears between $\beta_2$ and $\lr{\varepsilon_3^2}$ (see Figs.~\ref{fig:app2b} and \ref{fig:app3b}). The reason can be explained as follows. In large U+U collision system, one notes that the maximum influence of $\beta_3$ to $\lr{\varepsilon_2^2}$ appears at larger $\npart$ or $\nqp$ (around 3.5\% centrality) than the location of maximum influence of $\beta_2$ to $\lr{\varepsilon_3^2}$ (around 6\% centrality). In smaller Zr+Zr collision system, the peak location shifts towards more peripheral region, maximum influence of $\beta_3$ to $\lr{\varepsilon_2^2}$ shifts to around 7\% centrality and the maximum influence of $\beta_3$ to $\lr{\varepsilon_2^2}$ shifts to around 15\% centrality. In the latter case, the fluctuation-driven $\lr{\varepsilon_3^2}$ component is much more important than enhancement from $\beta_2$. In the former case, the $\lr{\varepsilon_2^2}$ value for undeformed case is also larger, leading to a smaller relative increase compare to U+U for the same $\beta_3$.

\begin{figure}[h!]
\begin{center}
\includegraphics[width=1\linewidth]{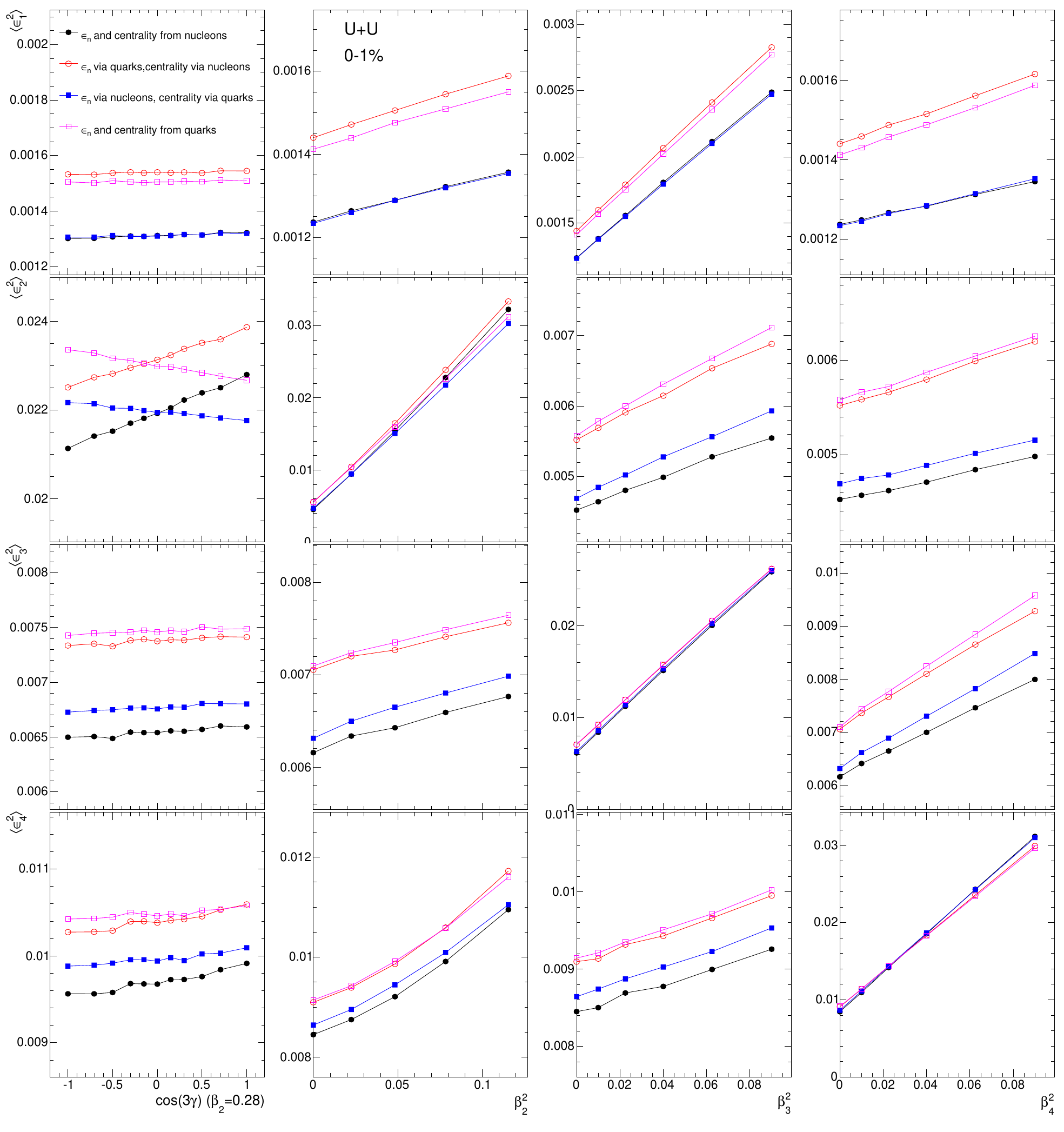}
\end{center}
\caption{\label{fig:app4}  The parametric dependence of $\lr{\varepsilon_n^2}$ in 0--1\% centrality U+U collisions on the quadrupole triaxiality $\gamma$ for $\beta_2=0.28$ (left column), axial quadrupole deformation $\beta_2$ (second column), axial octupole deformation $\beta_3$ (third column) and axial hexadecapole deformation $\beta_4$ (last column) for $n=1$ (top row), $n=2$ (second row), $n=3$ (third row) and $n=4$ (bottom row). The $\lr{\varepsilon_n^2}$ and centrality can be determined either from nucleon Glauber or quark Glauber, leading to four different curves in each panel as indicated by the legend in the top-left panel. They are obtained directly from plots like Figs.~\ref{fig:app0} and \ref{fig:app1}, where each panel provides the two set of data points in the corresponding panel in this figure.}
\end{figure}

\begin{figure}[h!]
\begin{center}
\includegraphics[width=1\linewidth]{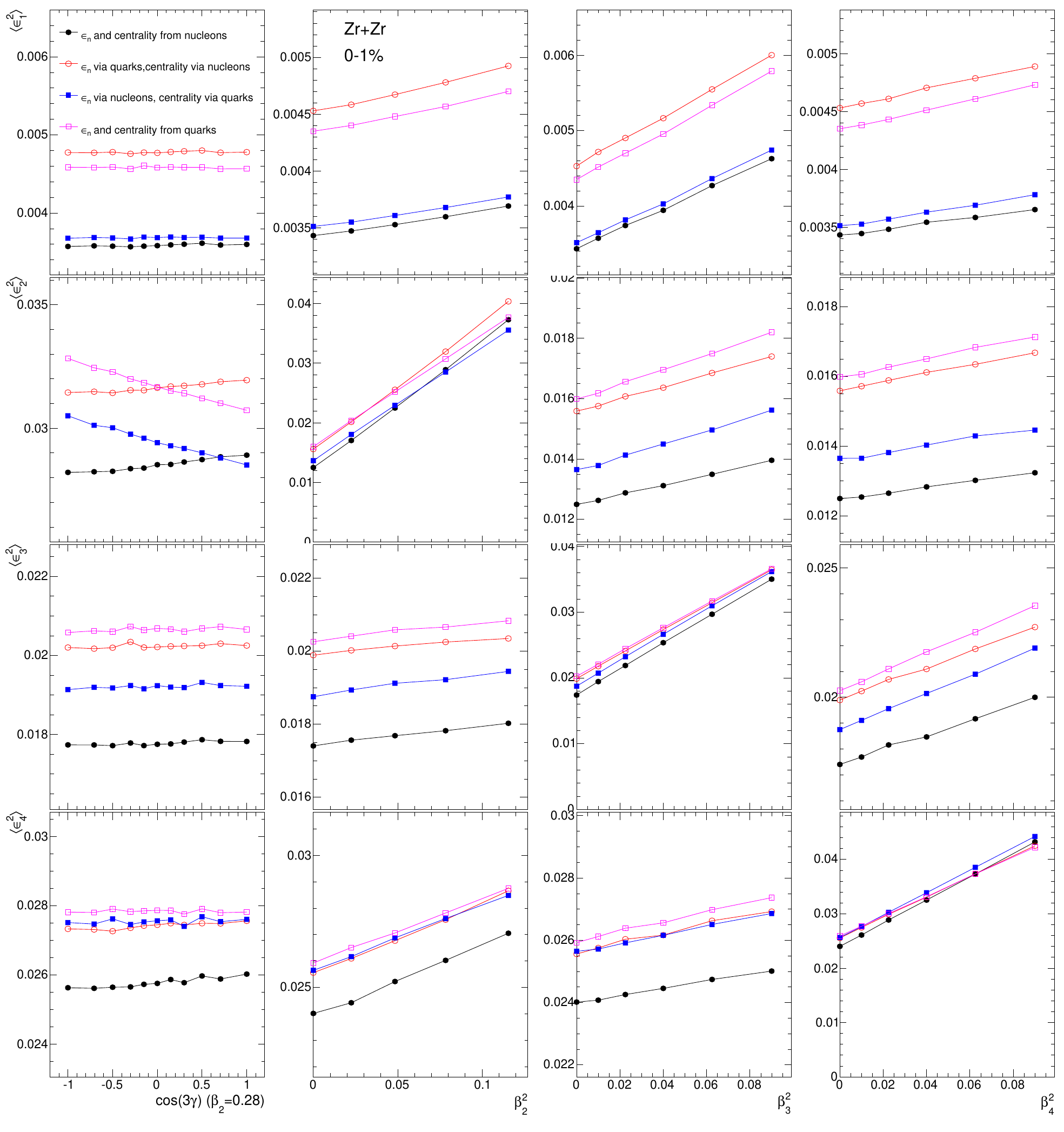}
\end{center}
\caption{\label{fig:app5}  Same as Fig.~\ref{fig:app4} but for 0--1\% most central Zr+Zr collisions.}
\end{figure}

Focusing on the 0--1\% centrality range, I then obtain the $\lr{\varepsilon_n^2}$ as a function of various $\beta_n^2$. The results are summarized in Figs.~\ref{fig:app4} and \ref{fig:app5} for U+U and Zr+Zr collisions, respectively. In most cases, strict linear dependencies are observed. One noticeable exception is the relation between $\lr{\varepsilon_4^2}$ and $\beta_2$, which is better described by a quartic dependence $\beta_2^4$ in the UCC region of U+U, consistent with the analytical results in Table~\ref{tab:2}. However, in the mid-central and peripheral U+U collisions and in Zr+Zr over the full centrality range, one finds that it is still better described by a $\beta_2^2$ dependence. Lastly, the slopes of these dependencies are nearly independent of whether nucleons or quarks are used for $\varepsilon_n$ or the centrality, with the exception of the $\gamma$ dependence of $\lr{\varepsilon_2^2}$.  

\begin{figure}[h!]
\begin{center}
\includegraphics[width=1\linewidth]{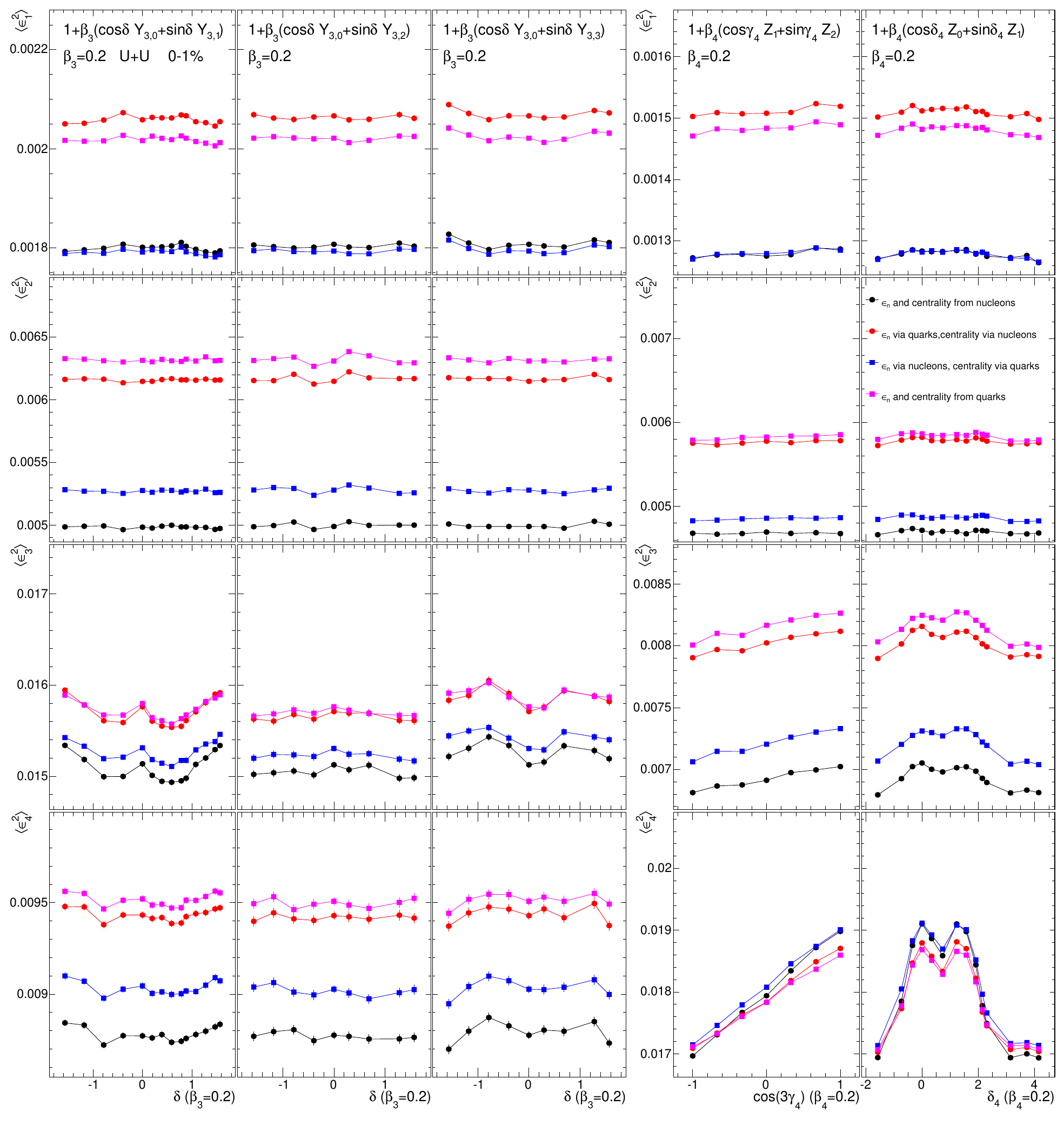}
\end{center}
\caption{\label{fig:app6} Effects of mixing between three pairs of octupole components in the left three columns, $1+\beta_3(\cos\delta Y_{3,0}+\sin\delta Y_{3,1})$, $1+\beta_3(\cos\delta Y_{3,0}+\sin\delta Y_{3,2})$ and $1+\beta_3(\cos\delta Y_{3,0}+\sin\delta Y_{3,3})$, and between two pairs of hexadecapole components in the right two columns, $1+\beta_4(\cos\gamma_4 Z_{1}+\sin\gamma_4Z_2)$ and $1+\beta_4(\cos\delta_4 Z_{0} + \sin\delta_4 Z_{1})$, where $Z_{0}=\sqrt{\frac{7}{12}}Y_{4,0}+ \sqrt{\frac{5}{12}}Y_{4,4}$, $Z_{1}=\sqrt{\frac{5}{12}}Y_{4,0}-\sqrt{\frac{7}{12}}Y_{4,4}$ and $Z_{2} = Y_{4,2}$ (see text). The results are obtained for 0--1\% central U+U collisions and are presented separately for $\lr{\varepsilon_n^2}$, $n=1$, 2, 3, and 4 from the top to the bottom rows. The $\lr{\varepsilon_n^2}$ and centrality are determined either from nucleon Glauber or quark Glauber, leading to four different curves in each panel as indicated by the legend in the right panel of the second row.}
\end{figure}

Section~\ref{sec:42} discusses briefly the influence of non-axial higher-order deformation, analogous to the triaxiality for the quadrupole deformation. This aspect is explored by mixing two different octupole or hexadecapole components, while keeping the overall magnitude of the deformation to be the same. Three cases are studied for the octupole deformation, $1+\beta_3(\cos\delta Y_{3,0}+\sin\delta Y_{3,1})$, $1+\beta_3(\cos\delta Y_{3,0}+\sin\delta Y_{3,2})$ and $1+\beta_3(\cos\delta Y_{3,0}+\sin\delta Y_{3,3})$, the results are shown in the left three columns of Fig.~\ref{fig:app6}. Only a small, less than 3\%, dependence on the mixing angle $\delta$ is observed for $\lr{\varepsilon_3^2}$. The situation for hexadecapole is a bit more involved. To simplify the discussion, I consider only the components respecting all three reflection symmetries, $Y_{4,0}$, $Y_{4,2}$ and $Y_{4,4}$. The nuclear surface can be parametrized with two angular variables $\gamma_4$ and $\delta_4$, in addition to $\beta_4$~\cite{Magierski:1996ic,Rohozinski:1997zz}:
\begin{align}\nonumber
R(\theta,\phi) = R_0(1+\beta_4(\cos\delta_4 Z_{0} + \sin\delta_4 [\cos\gamma_4 Z_{1}+\sin\gamma_4Z_2])),\;\\\label{eq:15}
Z_{0}= \sqrt{\frac{7}{12}}Y_{4,0}+ \sqrt{\frac{5}{12}}Y_{4,4},\; Z_{1}= \sqrt{\frac{5}{12}}Y_{4,0}-\sqrt{\frac{7}{12}}Y_{4,4},\; Z_{2} = Y_{4,2}
\end{align} 
The parameter $\gamma_4$ plays the similar role as the triaxiality parameter $\gamma$. For example for $\delta_4 = \mathrm{acos}(\sqrt{7/12})$,  $\gamma_4=0,2\pi/3,$ and $4\pi/3$ would correspond to axial-hexadecapole shape around $z$-, $x$- and $y$-axis, respectively. The right two columns of Fig.~\ref{fig:app6} show results for the two mixing cases, $1+\beta_4(\cos\gamma_4 Z_{1}+\sin\gamma_4Z_2)$ and $1+\beta_4(\cos\delta_4 Z_{0} + \sin\delta_4 Z_{1})$, respectively. A clear linear dependence on $\cos3\gamma_4$ is observed in the first case. The dependence in the second case is somewhat more complex, but I do observe it reaches maximum when $\delta_4=0$ or $\pi/2$, for which hexadecapole shape is described by  $Z_{0}=\sqrt{\frac{7}{12}}Y_{4,0}+ \sqrt{\frac{5}{12}}Y_{4,4}$ or $Z_{1}=\sqrt{\frac{5}{12}}Y_{4,0}-\sqrt{\frac{7}{12}}Y_{4,4}$ and positive $\beta_4=|\beta_4|$. The minimum on the other hand corresponds to the same shape components but with $\beta_4=-|\beta_4|$. This behavior is similar to the influence the prolate vs oblate quadrupole deformation on the $\lr{\varepsilon_2^2}$ as seen in Fig.~\ref{fig:3}.


\clearpage
\bibliography{deform2}{}
\bibliographystyle{apsrev4-1}

\end{document}